\RequirePackage{lineno} 
\documentclass[preprint2]{emulateapj}
\usepackage{amsmath}
\shorttitle{Rotational Light Curve of Jupiter}
\shortauthors{Ge et al.}
\usepackage{multirow}
\usepackage{textcomp}
\usepackage[colorlinks,
					citecolor=blue,
					linkcolor=blue]{hyperref}

\slugcomment{Accepted at AJ on January 1, 2019}
\begin{document}

\title{Rotational Light Curves of Jupiter from UV to Mid-Infrared and Implications for Brown Dwarfs and Exoplanets\\}

\author{$\rm Huazhi\ Ge^{1*}$, $\rm  Xi\ Zhang^1$, $\rm Leigh\ N.~Fletcher^2$, $\rm Glenn\ S.~Orton^3$, $\rm James\ Sinclair^3$, $\rm Josh\ Fernandes^3$, $\rm Tom\ Momary^3$, $\rm Yasumasa\ Kasaba^4$, $\rm Takao\ M.~Sato^5$, $\rm Takuya\ Fujiyoshi^6$}

\affil{
$^1$ Department of Earth and Planetary Sciences, University of California Santa Cruz,\\
 Santa Cruz, California 95064, USA\\
$^2$ Department of Physics \& Astronomy, University of Leicester,\\
 University Road, Leicester LE1 7RH, UK\\
$^3$ Jet Propulsion Laboratory, California Institute of Technology,\\
 4800 Oak Grove Drive, Pasadena, California 91109, USA\\
$^4$ Planetary Plasma and Atmospheric Research Center, Tohoku University,\\
 Aramaki-aza-Aoba 6-3, Aoba, Sendai, Miyagi 980-8578, Japan\\
$^5$ Institute of Space and Astronomical Science, Japan Aerospace Exploration Agency,\\
 3-1-1, Yoshinodai, Chuo-ko, Sagamihara, Kanagawa 252-5210, Japan\\
$^6$ Subaru Telescope, National Astronomical Observatory of Japan, National Institutes of Natural Sciences,\\
 650 N. A'ohoku Place, Hilo, HI 96720, USA}
\email{huazhige@ucsc.edu}

\begin{abstract}

Rotational modulations are observed on brown dwarfs and directly imaged exoplanets, but the underlying mechanism is not well understood. Here, we analyze Jupiter's rotational light curves at 12 wavelengths from the ultraviolet (UV) to the mid-infrared (mid-IR). Peak-to-peak amplitudes of Jupiter's light curves range from sub percent to 4\% at most wavelengths, but the amplitude exceeds 20\% at 5 $\rm \mu m$, a wavelength sensing Jupiter's deep troposphere. Jupiter's rotational modulations are primarily caused by discrete patterns in the cloudless belts instead of the cloudy zones. The light-curve amplitude is dominated by the sizes and brightness contrasts of the Great Red Spot (GRS), expansions of the North Equatorial Belt (NEB), patchy clouds in the North Temperate Belt (NTB) and a train of hot spots in the NEB. In reflection, the contrast is controlled by upper tropospheric and stratospheric hazes, clouds, and chromophores in the clouds. In thermal emission, the small rotational variability is caused by the spatial distribution of temperature and opacities of gas and aerosols; the large variation is caused by the $\rm NH_{3}$ cloud holes and thin-thick clouds. The methane-band light curves exhibit opposite-shape behavior compared with the UV and visible wavelengths, caused by wavelength-dependent brightness change of the GRS. Light-curve evolution is induced by periodic events in the belts and longitudinal drifting of the GRS and patchy clouds in the NTB. This study suggests several interesting mechanisms related to distributions of temperature, gas, hazes, and clouds for understanding the observed rotational modulations on brown dwarfs and exoplanets. 


\end{abstract}

\section{Introduction}
\label{sec:introduction}

	Rotational photometric variabilities have been observed on brown dwarfs (e.g., \citealt{radigan2012large}; \citealt{yang2016extrasolar}; \citealt{apai2017zones}), exoplanets (\citealt{zhou2016discovery}) and solar system gas giants (e.g., \citealt{gelino2000variability}; \citealt{karalidi2015aeolus}; \citealt{simon2016neptune}). The observed rotational variabilities range from sub-percent to tens of percent (e.g.,\citealt{artigau2009photometric}; \citealt{radigan2012large}; \citealt{heinze2013weather}; \citealt{metchev2015weather}). The observed light curves also show short-term and long-term evolution and wavelength-dependent signatures (e.g., \citealt{buenzli2012vertical}; \citealt{radigan2012large}; \citealt{yang2016extrasolar}; \citealt{apai2017zones}). These observations indicate that these substellar atmospheres possess complex meteorology that evolves with time.
	
	The current leading interpretation is that patchy clouds at different pressure levels cause the observed photometric variability (e.g., \citealt{radigan2012large, apai2013hst, apai2017zones}). Observations and theoretical models suggest that various types of clouds, such as silicates, salts, and metals, can form on brown dwarfs and close-in exoplanets (e.g., \citealt{ackerman2001precipitating, burgasser2002evidence, morley2012neglected, tan2016atmospheric, powell2018formation}). Recent studies suggest the existence of water clouds on one of the nearest type-Y brown dwarfs (WISE 0855), which has an effective temperature similar to Jupiter's water cloud level (\citealt{skemer2016first, morley2018band}). Previous studies have investigated the relationship between patchy cloud patterns and the photometric variability. The thin-thick cloud scenario proposed in \cite{apai2013hst} could explain the pressure-dependent bahaviours of multi-wavelength light curves. \cite{apai2017zones} suggested that long-term light-curve evolution on brown dwarfs is generated by beating patterns between planetary waves influencing the distribution of clouds. \cite{morley2014spectral} suggested that patchy clouds could generate large variability in the near-IR on objects warmer than 375 K. For objects cooler than 375 K, a larger-amplitude variability could be produced in the mid-IR due to water condensation. Alternatively, it was suggested that temperature fluctuations could also generate rotational photometric variabilities (e.g., \citealt{morley2014spectral, robinson2014temperature, zhang2014atmospheric}). Unfortunately, due to the difficulty of spatially resolving the atmospheres of remote brown dwarfs and exoplanets, the mechanisms behind the rotational photometric variability are still under debate.

\begin{deluxetable*}{ccccccccc}
\tablecolumns{7} 
\tablewidth{0pt}  
\tablecaption{\label{Table:Observation}Observational Data Information\\
\textsl{Limb-darkening coefficient at reflected sun-light and thermal emission wavelengths, Minnaert k, $\rm C_{0}$ and $\rm C_{1}$, are shown in the table}}

\bigskip

\tablehead{
\colhead {Central} &
\colhead {Filter} &
\colhead {Instrument} &
\colhead {Date} &
\colhead {Minnaert k} &
\colhead {$C_{0}$} &
\colhead {$C_{1}$} &
\colhead {Category} &
\colhead {Opacity} \\
\colhead{Wavelength} &
\colhead{Width} &
\colhead{Telescope} &
\colhead{} &
\colhead{} &
\colhead{} &
\colhead{} &
\colhead{} &
\colhead{} 
}
\startdata
0.275 $\rm \mu m$        &   &WFC3/HST       &Feb.  9, 2016     &0.520    &    &    & UV                  & Haze absorption \\ 
                                      &   &                         &Apr.  3, 2017      &0.520    &    &    &                        &  \\
\\
0.343 $\rm \mu m$        &   &WFC3/HST     &Jan. 19, 2015      &0.850    &    &    & blue               & Chromophores\\
                                      &   &                       &Apr.  3, 2017       &0.850    &    &    &                       & \\
\\
0.395 $\rm \mu m$        &   &WFC3/HST     &Jan. 19, 2015    &0.850    &    &    & blue                 & Chroophores\\
                                      &   &                        &Feb.  9, 2016    &0.850    &    &    &                         & \\
                                      &   &                        &Apr.  3, 2017    &0.850     &    &    &                          &\\
\\
0.467 $\rm \mu m$        &  &WFC3/HST     &Feb.  9, 2016     &0.950    &    &    & blue                & Chromophores\\
                                      &  &                        &Apr.  3, 2017     &0.950    &    &    &                         & \\
\\
0.502 $\rm \mu m$        &  &WFC3/HST      &Jan. 19, 2015     &0.950    &    &    & green            & Chromophores\\
                                      &  &                         &Feb.  9, 2016     &0.950    &    &    &                       & \\
                                      &  &                         &Apr.  3, 2017      &0.950    &    &    &                       &\\
\\
0.547 $\rm \mu m$        &  &WFC3/HST      &Feb.  9, 2016      &0.970    &    &    & green            & Chromophores\\

\\
0.631 $\rm \mu m$        &  &WFC3/HST       &Jan. 19, 2015     &0.999    &    &    & red               & Chromophores\\
                                      &  &                         &Feb.  9, 2016     &0.999    &    &    &                      &\\
                                      &  &                         &Apr.  3, 2017      &0.999    &    &    &                      & \\
\\
0.658 $\rm \mu m$        &  &WFC3/HST       &Jan. 19, 2015       &0.999    &    &    & red             & Chromophores\\
                                      &  &                          &Feb.  9, 2016       &0.999    &    &    &                    &  \\
                                      &  &                          &Apr.  3, 2017        &0.999    &    &    &                    &\\
\\
0.889 $\rm \mu m$        &  &WFC3/HST       &Jan. 19, 2015    &1.000    &    &    & near-IR          & $\rm CH_{4}$\\
                                      &  &                          &Feb.  9, 2016    &1.000    &    &    &                        &\\
                                      &  &                          &Apr.  3, 2017     &1.000    &    &    &                        &\\
\\
5.1 $\rm \mu m$            &0.25 $\rm \mu m$  &Spex/IRTF            & May. 11-12, 2016  &    &    &    &mid-IR       & Atmosphere Window\\
\\\
8.59 $\rm \mu m$          &0.42 $\rm \mu m$  &VISIR/VLT             & Feb. 15-16, 2016  &    &0.032  &0.120    & mid-IR      & $\rm PH_{3}$  \\
\\
8.80 $\rm \mu m$          &0.8 $\rm \mu m$  &COMICS/Subaru   & Jan. 24-25, 2016 &    &0.083  &0.119    & mid-IR      & $\rm PH_{3}$  \\

\\
10.50 $\rm \mu m$        &1.0 $\rm \mu m$   &COMICS/Subaru   & Jan. 24-25, 2016 &    &0.152  &0.202    & mid-IR      & $\rm NH_{3}$  \\
\\
10.77 $\rm \mu m$        &0.19 $\rm \mu m$   &VISIR/VLT             & Feb. 15-16, 2016  &    &0.102  &0.160    & mid-IR      & $\rm NH_{3}$  

\enddata

\end{deluxetable*}

With spatially resolved maps, gas giant atmospheres in the solar system shed lights on the mechanisms behind rotational light curves. Previous studies indicated that Jupiter and Neptune exhibit rotational modulations (\citealt{gelino2000variability}; \citealt{karalidi2015aeolus}; \citealt{simon2016neptune}; \citealt{stauffer2016spitzer}). The magnitudes of the photometric variabilities are also wavelength-dependent, ranging from sub-percent level to $\sim$10\% at different wavelengths on Jupiter and Neptune. The atmosphere patterns can be retrieved from light-curve evolution. For instance, \cite{karalidi2015aeolus} provided a mapping tool to retrieve the Great Red Spot and 5-$\rm \mu m$ hot spots on Jupiter from the light curves of two consecutive rotations at the UV and the near-IR. \cite{simon2016neptune} retrieved the jet speeds on Neptune using the power spectra of 50-day light curves at the visible wavelengths from Kepler observation.

However, most previous studies on rotational modulations of Jupiter and Neptune focused on the wavelengths dominated by sun-light reflection (\citealt{gelino2000variability}; \citealt{karalidi2015aeolus}; \citealt{simon2016neptune}; \citealt{stauffer2016spitzer}). The thermal-emission light curves have not been thoroughly investigated yet. Because the light curves of free-floating brown dwarfs are observed in thermal emission, the infrared light curves of Jupiter might provide a better insight to understand the observed rotational variability of exoplanets and brown dwarfs. Furthermore, weather patterns on Jupiter might be similar to the cold Y-type brown dwarfs (effective temperature $\sim$300 to 500 K, e.g., \citealt{skemer2016first}; \citealt{morley2018band}), which might exhibit zonally banded patterns and storms (\citealt{zhang2014atmospheric}).

\begin{figure*}
\centering 
\includegraphics[width = 1.0\textwidth]{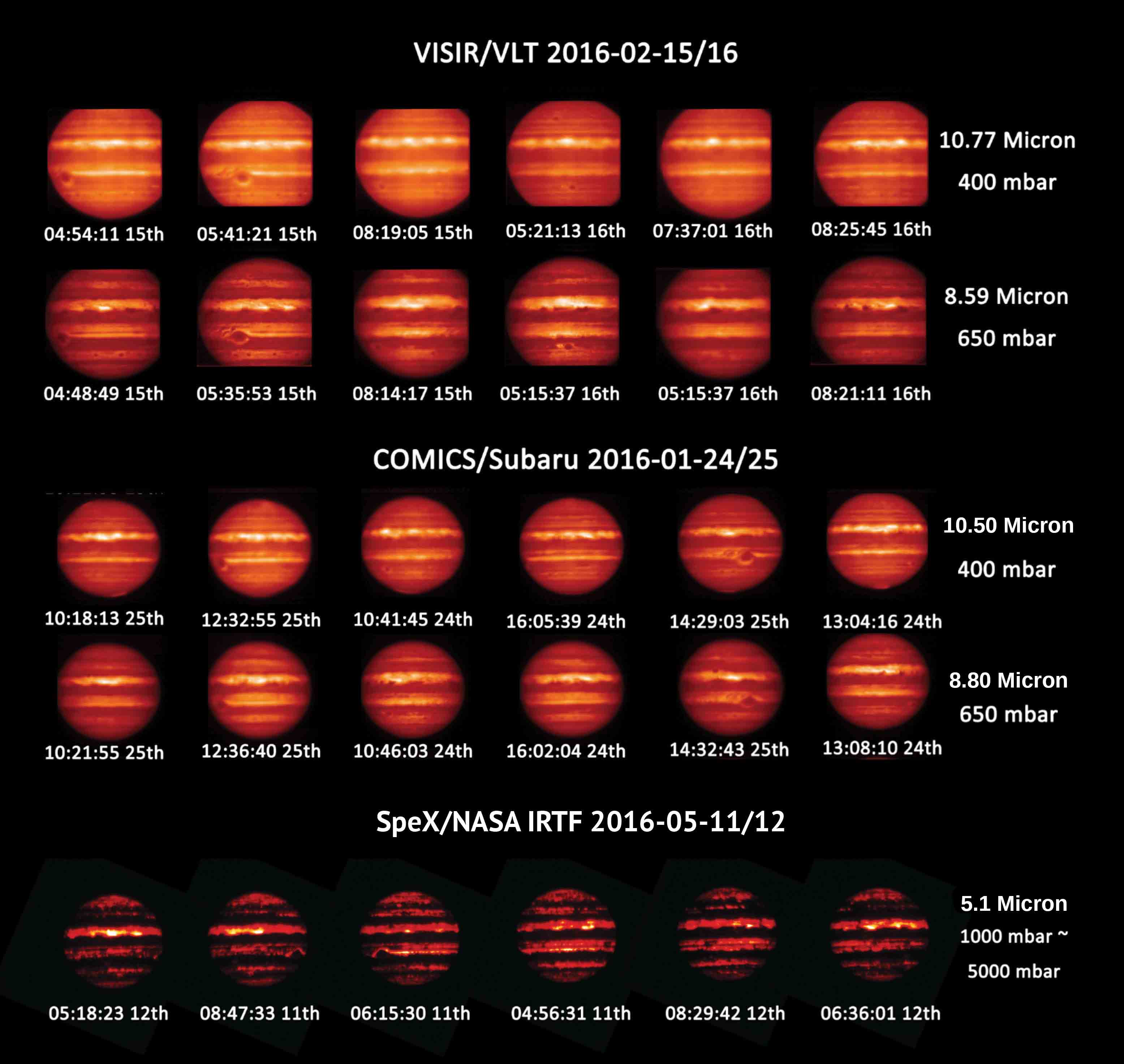} 
\caption{\label{Fig:DiskList} Images of Jupiter obtained at different time and wavelengths from VISIR/VLT (top), COMICS/Subaru (middle) and SpeX/IRTF (bottom). These images are combined to generate global cylindrical radiance maps in this study after limb-darkening correction.}
\end{figure*}

Jupiter's thermal-emission observations have provided high spatial resolution maps of temperature and chemical compositions ($\rm NH_{3}$, $\rm PH_{3}$, hydrocarbons, and aerosols) at multiple pressure levels from $\rm NH_{3}$ clouds condensation level to the stratosphere (e.g., \citealt{orton1991thermal, simon2016neptune, fletcher2016mid}). Following the convention, in this paper we use ``cloud'' to represent the condensed volatiles such as $\rm NH_{3}$, $\rm NH_{4}HS$ and $\rm H_{2}O$. ``Chromophore'' stands for the color agents (such as the reddish species) in the GRS and the other regions in and above the $\rm NH_{3}$ clouds (\citealt{carlson2016chromophores}; \citealt{sromovsky2017possibly}). We use ``haze'' to stand for the particles above and close to the main $\rm NH_{3}$ cloud top, including upper tropospheric and stratospheric aerosol particles (i.e., \citealt{west1991evidence}; \citealt{west2004jovian}; \citealt{zhang2013stratospheric, fletcher2016mid}). The term ``aerosol'' is a general name for all particles including hazes. The distributions of deep $\rm H_{2}O$ clouds (8 to 5 bars), $\rm NH_{4}SH$ (2 to 1.5 bars) and $\rm NH_{3}$ clouds (0.7 to 0.3 bars) in the troposphere \citep{west2004jovian}, hazes in the upper troposphere and stratosphere, along with chromophores in and above the $\rm NH_{3}$ clouds, have all been well observed and studied for decades (e.g., \citealt{ferris1987hcn, west2004jovian, taylor2004composition, carlson2016chromophores, sromovsky2017possibly}). These data allow us to thoroughly investigate the underlying mechanism of rotational light curves of Jupiter in thermal emission and quantify the important roles of temperature, gas and clouds.
	
In this study, we analyze Jupiter's light curves at both reflection and thermal-emission wavelengths and provide interpretations on what controls the light-curve amplitudes and wavelength-dependent behaviors. We construct the sun-light reflected light curves from the radiance maps imaged by WFC3/UVIS on the Hubble Space Telescope (HST) from the UV to the near-IR wavelengths in nine channels. The thermal-emission light curves are constructed from images obtained by the COMICS instrument on the Subaru Telescope \citep{kataza2000comics}, the VISIR instrument on the VLT \citep{lagage2004successful}, and SpeX instrument on NASA's Infrared Telescope Facility (IRTF) in three mid-IR channels. We describe the data acquisition in Section~\ref{sec:data}  and light curve construction in Section~\ref{sec:mosaick map and light curve construction}. In Section~\ref{sec:mosaick map and light curve construction}, we also introduce a new method to quantitatively estimate the contribution of the rotational modulation from each latitude to identify the locations of critical discrete patterns. In Section~\ref{sec:Jupiter maps} we overview the important discrete patterns and periodic events that are responsible for Jupiter's rotational modulation. Section~\ref{sec:Jupiter Light Curve} discusses how the light-curve amplitudes, shapes and evolution are influenced by the sizes, brightness, locations and evolution of discrete patterns. We will also discuss the underlying physical and chemical mechanisms. Section~\ref{sec:implications} describes the implications of this study on the multi-wavelength rotational variabilities on brown dwarfs and directly imaged exoplanets. We conclude this study in Section~\ref{sec:conclusions}.

\section{Data}
\label{sec:data}

\subsection{Reflection Data}
\label{sec:reflection data}

We use the global maps of Jupiter in reflected sunlight obtained during the Outer Planets Atmosphere Legacy (OPAL) program by the Hubble Space Telescope's (HST) Wide Field Camera 3 (WFC3, \citealt{simon2015first}). The OPAL data cover a large wavelength range from the UV to the near-IR in 9 channels. The images at each wavelength are reduced via the standard Hubble processing pipeline (Section 2.2 in \citealt{simon2015first}). Jupiter is mapped from the latitude of $\rm -79.8\arcdeg$S to $\rm +79.8\arcdeg$N with a resolution of $\rm \sim$10 pixels/deg (corresponding to $\rm \sim$120 km at equator to $\rm \sim$600 km at the high latitudes). Information of the wavelengths, filters, observational dates and main opacity sources is listed in Table~\ref{Table:Observation}. More detailed information about the data is provided on the official OPAL website (e.g., \citealt{simon2015first}) \href{https://archive.stsci.edu/prepds/opal/}{https://archive.stsci.edu/prepds/opal/}.

\begin{figure}
\centering \includegraphics[width = 0.4\textwidth]{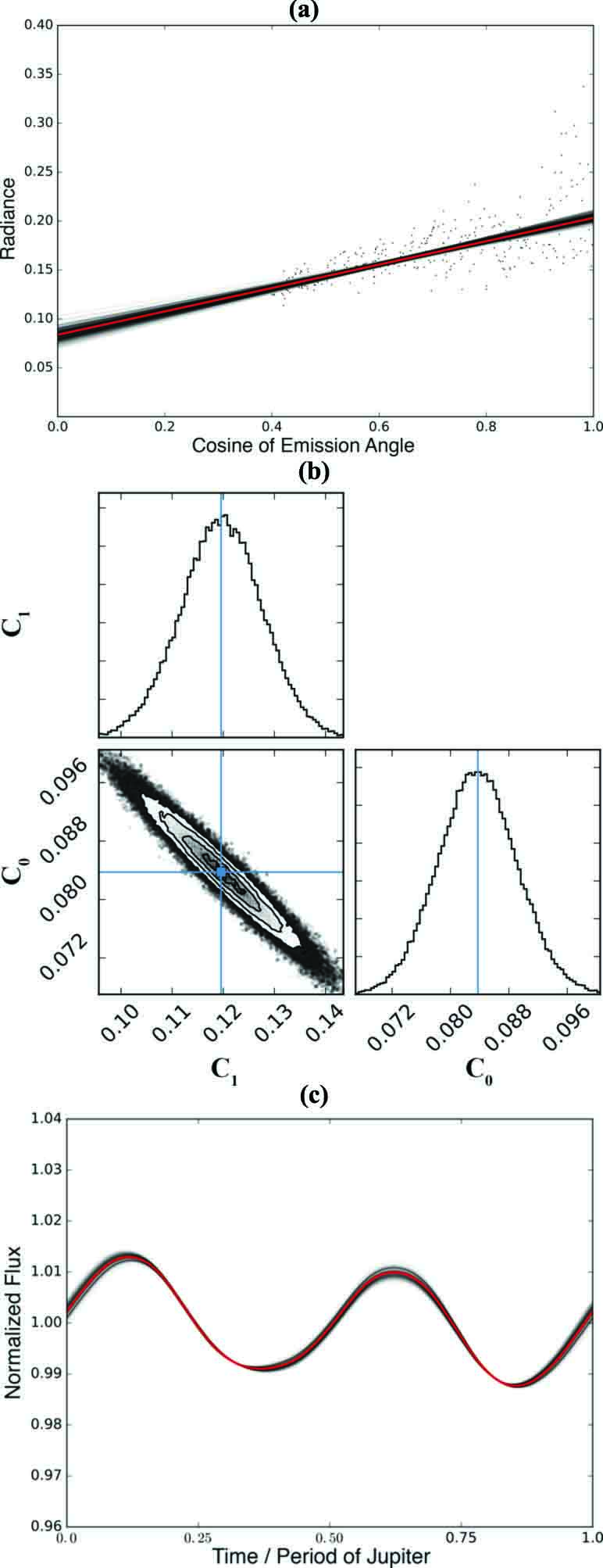} 
\caption{\label{Fig:LimbDark} An example of limb-darkening fitting at 8.80 $\rm \mu m$. (a) Limb darkening profile. The dotted points are randomly selected 300 pixels from all latitudes with cosine of the emission angles larger than 0.4. The red line denotes the fitting with $\rm 50^{th}$ percentile of the samples in the marginal distribution. (b) The posteriors of the fitting coefficients. (c) The uncertainty of the light curve shape at 8.80 $\rm \mu m$. The red curve is constructed by the red fitted line in (a). Black curves are constructed by the black fitted lines in (a).}
\end{figure}

Since 2014, OPAL program has provided Jupiter's multi-wavelength maps of two Jupiter's rotations in each year. But Jupiter's moons (e.g., Io) cast shadows on some maps (e.g., all maps in Rotation B of Cycle 23). Also, some maps possess residual-fringe features (e.g., 889 nm map in Rotation A of Cycle 22). The major satellite shadows and residual fringes generate photometric variabilities which we are not interested in. Thus we excluded these OPAL maps. We chose the rotation B from cycle 22 at 02:00 to 12:30 UT on January 19, 2015, rotation A from cycle 23 at 09:35 to 18:04 UT on February 9, 2016, and rotation B from cycle 24 in April 2017 from the OPAL database. The one-year interval between these maps allows us to investigate long-term light-curve evolution on Jupiter. 

The observations in reflected sun-light at different wavelengths are sensitive to different pressure layers and compositions in Jupiter's atmosphere in and above the $\rm NH_{3}$ cloud layer. The UV (0.275 $\rm \mu$m) maps reveal the distributions of upper tropospheric hazes at equator and mid-latitude regions and stratospheric hazes at high latitudes (\citealt{west2004jovian, zhang2013stratospheric}). At the equator and the mid-latitude regions, the UV observation probes into the troposphere to the hundred-millibar level \citep{vincent2000mapping}. Due to abundant stratospheric hazes and the limb-darkening effects, the effective pressure layer (where optical depth $\rm \sim$1) is about tens of millibars in the high-latitude region. The visible-light observations (0.343 $\rm \mu$m, 0.395 $\rm \mu$m, 0.467 $\rm \mu$m, 0.502 $\rm \mu$m, 0.547 $\rm \mu$m, 0.631 $\rm \mu$m and 0.658 $\rm \mu$m) are also sensitive to the chromophores in the $\rm NH_{3}$ clouds, probing pressure levels of a few-hundred millibars. In the strong methane band at 0.889 $\rm \mu$m, the observations would probe the 600-mbar pressure level in the absence of clouds \citep{west2004jovian}. However, in the presence of patchy and discrete clouds, the effective pressure levels are much higher in altitude, being strongly affected by the cloud and haze opacities.

\subsection{Thermal-Emission Data}
\label{sec:thermal emission data}

Thermal-emission snapshots of Jupiter's maps were provided by the VISIR instrument \citep{lagage2004successful} on the Very Large Telescope (VLT) in February 2016 (program ID 096.C-0091), the COMICS instrument \citep{kataza2000comics} on the Subaru Telescope in January 2016 (program ID S16B-049), and the SpeX instrument on NASA's Infrared Telescope Facility (IRTF) in May 2016 (program ID 2016A-022). Both VLT and Subaru have about 8-m diameter primary mirrors, providing sufficient spatial resolution to resolve discrete features on Jupiter's disc. Both COMICS and VISIR observe Jupiter in multiple wavelengths from 7-25 $\rm \mu$m---we have selected wavelengths near 8.7-8.8 $\rm \mu$m (sensing $\rm \sim$600-mbar tropospheric $\rm PH_{3}$, aerosol opacity and temperature) and 10.3-10.7 $\rm \mu$m (sensing tropospheric ammonia and temperature near 400-500 mbar) for this study. These are supplemented by observations at 5 $\rm \mu$m sensing Jupiter's mid-tropospheric emission, with condensation clouds in the 1-4 bar range appearing in silhouette against this bright thermal background. According to some recent studies (e.g., \citealt{bjoraker2015jupiter}), it was suggested that the observed pressure broadening of lines at 5 $\rm \mu m$ implies the presence of a water cloud, but the narrow-band imaging used in this study blend together the contributions from clouds and gases to provide a broad and extended contribution function over 1-to-4 bar range. The SpeX/IRTF observations use the smaller 3-m primary mirror. Figure~\ref{Fig:DiskList} shows these three different wavelengths; for each wavelength we have six disk images acquired through two consecutive nights. The spatial resolution varies from $\rm 0.2\arcsec$ to $\rm 0.4\arcsec$ ($\rm \sim$700 to $\rm \sim$1400 km on Jupiter). Although the thermal-emission maps of Jupiter have coarser resolution than the reflection maps, the spatial resolution is good enough to resolve most small-scale discrete patterns, such as ovals in the southern hemisphere and 5-$\rm \mu$m hot spots (Figure~\ref{Fig:DiskList}). We use a similar data reduction procedure as described by \cite{fletcher2009retrievals} and \cite{fletcher2016mid}, for the bad-pixel removal, flat fielding, limb fitting and cylindrical map projection. Observations at 8.8 and 10.5 $\rm \mu$m were radiometrically calibrated to match results from Cassini's Composite Infrared Spectrometer (CIRS) during its Jupiter flyby in 2000 \citep{fletcher2016mid}. We note that Jupiter observations with VISIR over-fill the $\rm 38\times38\arcsec$ field of view, meaning that one of the poles is always omitted from the observation in Figure~\ref{Fig:DiskList}, whereas the field of view of COMICS ($\rm 42\times32\arcsec$) is sufficient to capture the whole disc of Jupiter in a two-point mosaic. Taken together, these three wavelengths are sensitive to the distributions of temperature, gases (e.g., $\rm NH_{3}$, $\rm PH_{3}$ and $\rm CH_{3}D$), hazes, and clouds in the mid and upper troposphere of Jupiter. 

\section{Mosaicked Map and Light Curve Construction}
\label{sec:mosaick map and light curve construction}

Following the methods in \cite{gelino2000variability} and \cite{cowan2008inverting}, we construct Jupiter's light curves from Jupiter's cylindrical global maps using carefully characterized wavelength-dependent limb-darkening properties. 

\subsection{Reflection Light Curve Construction}
\label{sec:light curve construction}

At wavelengths sensing reflected sunlight, the OPAL dataset provides the cylindrical global maps as well as the limb-darkening formula and coefficients. The reflection light curves can be directly constructed via moving a simulated aperture scanning through the global radiance maps. The aperture is $\rm 180\arcdeg$-longitude wide and covers all latitudes. As the planet rotates in the aperture, we apply the limb-darkening effect on the radiance map. The light-curve flux is calculated by integrating the pixel radiance weighted by its projected area on a disk within the aperture:

\begin{equation} \label{eq1}
f(\phi, t) = \int_{\Omega t}^{\Omega t+\pi} I(\lambda,\phi)L(\mu, \mu_{0}) \cos \lambda\ d\lambda,
\end{equation}

\begin{equation} \label{eq2}
F(t) = \int_{-\frac{\pi}{2}}^{\frac{\pi}{2}} f(\phi, t) \, {\cos^2 \phi} \, d\phi,
\end{equation}

\begin{equation} \label{eq3}
L(\mu, \mu_{0}) = I_{0}\cdot(\mu \mu_{0})^{k}/{\mu},
\end{equation}
where $f(\phi,t)$ is the rotational modulation of the flux at each latitude at time $t$; $\Omega$ is the rotation rate of Jupiter; $\lambda$ is the longitude; $\phi$ is the latitude; $F(t)$ is the disk-integrated flux. In the reflection limb-darkening formula, $L(\mu,\mu_0)$, $\mu$ is cosine of the emission angle; $\mu_0$ is cosine of the sunlight incident angle, which is fixed to be 1 (incident angle is $\rm 0\arcdeg$) in our light curve construction; $k$ is the limb-darkening coefficient for each map from the OPAL dataset (Table~\ref{Table:Observation}).

\subsection{Thermal-Emission Light Curve Construction}
\label{sec:mosaic construction}

\begin{figure*}
\centering \includegraphics[width = 1.0\textwidth]{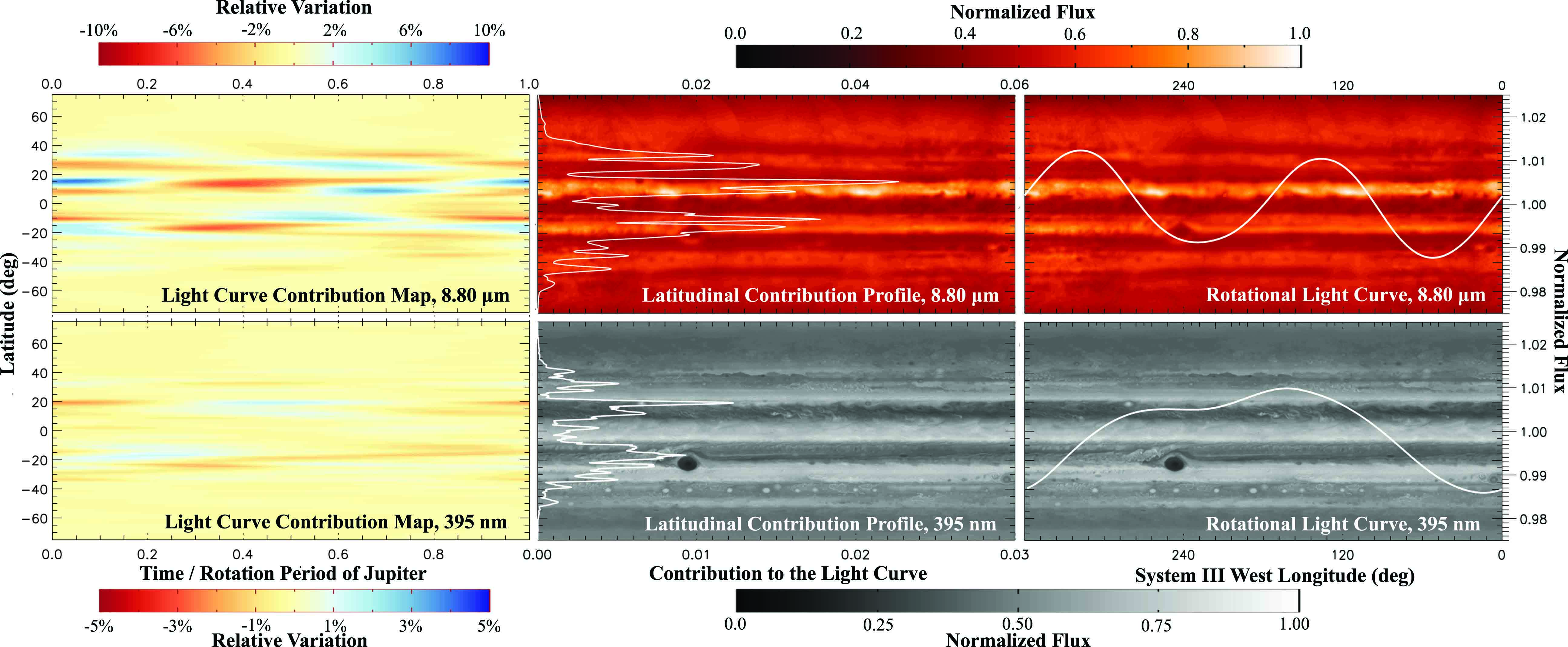} 
\caption{\label{Fig:Contribution Maps}Light-curve contribution maps (left column), maps with latitudinal contribution profiles (middle column), and global radiance map with rotational light curves (right column) at 395 nm (lower row) and 8.80 $\rm \mu m$ (upper row). The light-curve contribution maps and latitudinal contribution profiles indicate that the rotational variation is primarily exhibited in the belts and interfaces between the neighboring belts and zones. The light-curve-contribution profiles are also correlated with the global radiance map. The variations (e.g., red region and blue region), which are induced by the GRS and the NEB expansion events, can be identified on the light-curve contribution maps.}
\end{figure*}

In order to construct the thermal-emission light curves, we construct the cylindrical global maps without limb darkening in the following steps. First, we use the raw data reduction method presented in \cite{fletcher2009retrievals} to reduce the data and assign latitude, longitude and emission angle to each pixel. Second, we fit limb darkening in the mosaics at different wavelengths, using the limb-darkening formula:

\begin{equation} \label{eq4}
L(\mu) = (c_0 + c_1\mu)/(c_0 + c_1),
\end{equation}
where $\mu$ is cosine of the emission angle; $c_0$ is the zero-order coefficient and $c_1$ is the first-order coefficient. We use the Maximum Likelihood Estimation (MLE) method to fit the limb darkening. Because the radiances of the pixels with larger emission angles (i.e., smaller $\rm \mu$) usually have larger observational uncertainties, only the pixels with $\rm \mu$ larger than 0.4 are selected in the limb-darkening fitting. We split $\rm \mu$ from 0.4 to 1 into 6 uniform intervals with the interval width of 0.1, and randomly pick 50 pixels from each interval from $\rm -70\arcdeg$S to $\rm +70\arcdeg$N. Then we fit the limb-darkening formula using a Monte-Carlo Markov Chain package, EMCEE \citep{foreman2013emcee}. The free parameters are limb darkening coefficients, $c_0$ and $c_1$. The MLE equation is given by:

\begin{equation}
\begin{aligned}
\ln \, p(L_{0} &\mid \mu,  c_{0}, c_{1}, f_{1}, f_{2}) = \\
&-\frac{1}{2} \sum^{n} [\frac{(L_{0}(\mu_{n})-c_{1}\mu_{n}-c_{0})^2}{s_{n}^{2}} + \ln(2 \pi s_{n}^{2})]
\end{aligned}
\end{equation}

\begin{equation}
s_{n}^{2} = f_{1}^2 L_{0}(\mu_{n})^2 + f_{2}^2 (c_{1}\mu_{n} + c_{0})^2
\end{equation}
where $L_{0}(\mu_{n})$ is the pixel radiance; $\mu_{n}$ is cosine of the emission angle of a selected pixel. According to the differences between the calibration of IRIS and CIRS on Jupiter (calibration uncertainty is about 5-10\%), the systematic uncertainties of the radiance, $f_1$ and $f_2$, are both chosen to be 10\% in this study. We allow the term $f_{2}^2 (c_{1}\mu_{n} + c_{0})^2$ changing in the MCMC fitting. An example of limb-darkening fitting at 8.80 $\rm \mu m$ is shown in Figure~\ref{Fig:LimbDark}. Adopting a larger pixel uncertainty value (e.g., 20\% to 50\%) does not change the light curves; only the uncertainty of the resulting light curves (shaded region in the lower panel of Figure~\ref{Fig:LimbDark}) becomes larger. But it does not alter our conclusions in this study. We also tested fitting the limb darkening of belts and zones separately but the constructed light curves do not change too much either.

For each wavelength, we combine the individual cylindrical maps obtained from disk images (Figure~\ref{Fig:DiskList}) into a global radiance map and then we remove the limb-darkening effect from each pixel. We have to deal with the overlapping area where the two or more individual maps overlap. The traditional way is to first remove the limb darkening for each individual map and then average the data in the overlapping area (e.g., \citealt{fisher2016organization}). But we note that pixel radiances with smaller $\rm \mu$ usually have larger observational uncertainties as well as larger cylindrical projection errors. To reduce the errors, in this study we first combine the individual maps without removing the limb-darkening. In the overlapping area, we select the pixels with maximum radiances.\footnote {We also tried another metric and selected the pixels with maximum $\rm \mu$, but the two do not make big difference because the pixel with a larger $\rm \mu$ usually has a larger radiance.} Then we remove the limb-darkening effects of the final combined cylindrical maps to obtain a global radiance map in Figure~\ref{Fig:Maps}.

Lastly, we construct the thermal-emission light curves from the global radiance maps using Equations~\ref{eq1},~\ref{eq2}, and~\ref{eq4} to reintroduce the thermal-emission limb darkening $L( \mu)$ at every longitude. The limb-darkening removal and reintroduction could introduce uncertainties to the light curves. We quantify the uncertainty of the light-curve shape in Figure~\ref{Fig:LimbDark}c. The uncertainty of limb-darkening fitting influences the light-curve shape, but the effect is generally small.

\subsection{Contribution from Each Latitude}
\label{sec:latitudinal contribution}

We also develop a method to identify the locations of discrete patterns that primarily cause the rotational variability. First, instead of calculating the light curve for the entire globe, we construct the light curve at each latitude using Equation~\ref{eq1}. Then, we define the contribution of latitude $\phi$ at time $t$ to the global light curve as:

\begin{equation} \label{eq5}
C(\phi, t) = f(\phi, t) \cos^2{\phi} /\sigma[F(t)],
\end{equation}
where $\sigma[F(t)]$ is the standard deviation of global light curve, which represents the mean amplitude of the photometric variability.

We then construct the light-curve contribution maps and latitudinal contribution profiles from the latitude $\rm -75\arcdeg$S to $\rm +75\arcdeg$N (left and middle panels in Figure~\ref{Fig:Contribution Maps}). Note that the positive and negative contributions from different latitudes could cancel out each other. Then, we calculate the standard deviation of $C(\phi, t)$ of each latitude, which can be considered as the mean contribution of latitude $\phi$ to the amplitudes of global light curve. The latitudinal contribution profiles are shown in the middle panels of Figure~\ref{Fig:Contribution Maps} and left panels of Figure~\ref{Fig:Maps}.
	
The major discrete patterns induce large rotational variabilities in the belts in the light-curve contribution map (Figure~\ref{Fig:Contribution Maps}). For example, one can see the rotational variations caused by the GRS, NEB expansion event, and hot spots (i.e., blue regions caused by the bright patterns, red regions caused by the dark patterns) in the light-curve contribution map (Figure~\ref{Fig:Contribution Maps}). Note that this method is only applicable to the equator-on-observed atmospheres because we do not include any information of inclination angles here. Next, we will describe these discrete patterns in detail.

\section{Discrete Patterns on Jupiter's Reflection and Thermal-Emission Maps}
\label{sec:Jupiter maps}

\begin{figure*}
\centering \includegraphics[width = 0.85\textwidth]{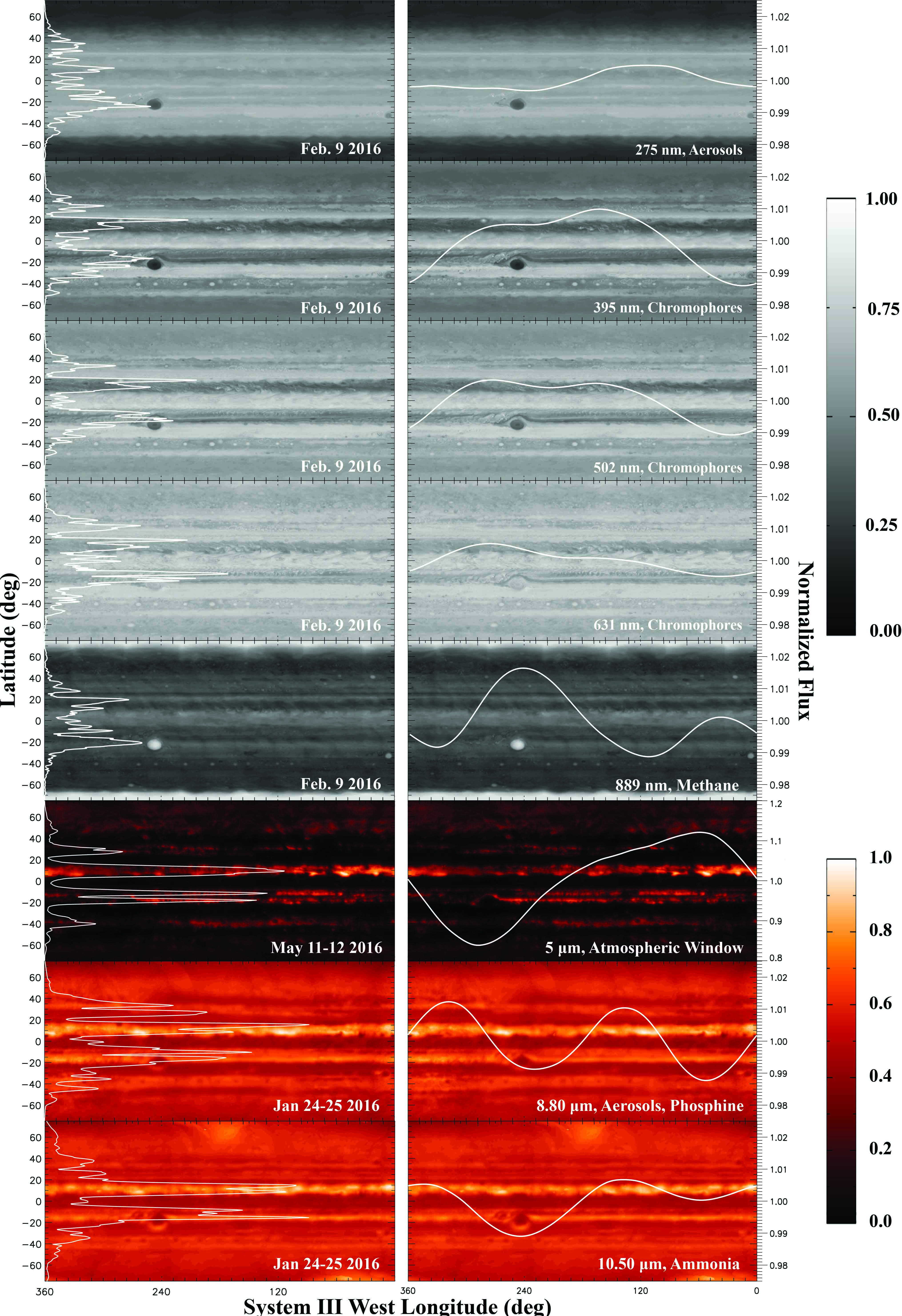} 
\caption{\label{Fig:Maps}Jupiter's global radiance maps from the UV to the mid-IR wavelengths, latitudinal contribution profiles (white lines in the left column), and rotational light curves (white lines in the right column). The latitudinal contribution profiles indicate that the belts exhibit more rotational modulations than the zones. The reflection latitudinal contribution profiles range from 0 to 0.03; the thermal emission profiles ranges from 0 to 0.06. }
\end{figure*}

Jupiter's brightness maps from the UV (275 nm) to the mid-IR (10.50 $\rm \mu m$) are presented in Figure~\ref{Fig:Maps}. Left panels show the cylindrical global maps with the latitudinal contribution profiles. Right panels are the same global radiance maps displayed with the corresponding light curves. Jupiter's global maps appear to vary significantly at different wavelengths because observations at different wavelengths probe into different pressure levels and they are sensitive to different weather patterns. For example, Jupiter's banded structure is not evident on the UV (275 nm) map and on the methane-band (889 nm) map, but belts and zones are distinct at visible and thermal-emission wavelengths. For another example, the GRS is a dark spot in the visible-wavelength maps but is a bright spot in the methane-band (889 nm) maps. These differences are controlled by distributions of temperature, chemical compositions, hazes, and clouds in the troposphere and stratosphere.

\begin{figure*}
\centering \includegraphics[width = 1.0\textwidth]{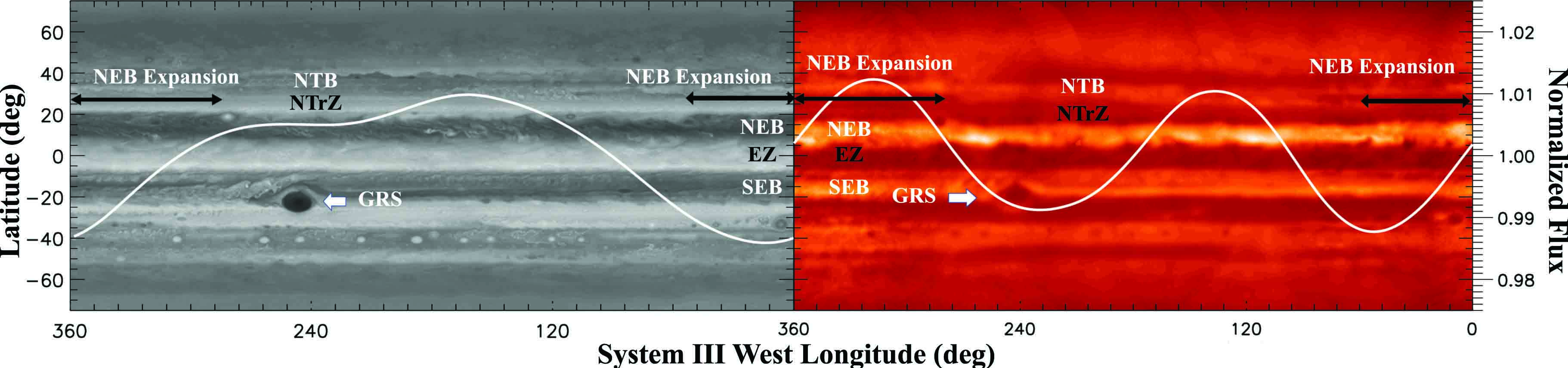} 
\caption{\label{Fig:Pattern Maps} Locations of dominant patterns, periodic events, belts and zones on typical Jupiter maps. The left panel is a reflected sun-light map imaged by WFC3/HST. The right panel is a thermal-emission map at 8.8 $\rm \mu m$ imaged by COMICS/Subaru. The rotational light curve is also shown in each plot to illustrate the correlation between discrete patterns and the light-curve shape. Note that the peaks and troughs of light curves do not precisely coincide with the location of the GRS and the NEB expansion event. }
\end{figure*}

Jupiter's atmosphere is generally dominated by the banded structure known as belts and zones. Belts are dark regions in reflected sunlight but appear bright and warm in thermal emission (Figure~\ref{Fig:Pattern Maps}). They are cloudless, hot, and dry regions exhibiting several periodic events, such as the North Equatorial Belt (NEB) expansion events, and fades and revivals of the South Equatorial Belt (SEB) (e.g., \citealt{ingersoll2004dynamics, fletcher2017moist, fletcher2017jupiter}). Zones are bright regions in reflected sunlight but appear dark in thermal emission (Figure~\ref{Fig:Pattern Maps}). They are cloudy, cold, moist and nearly longitudinally uniform. Belts and zones play different roles in the rotational modulations of Jupiter. The latitudinal contribution profiles (left panels in Figure~\ref{Fig:Maps}) indicate that belts exhibit much larger rotational modulations than zones at all wavelengths.

There are many discrete patterns in Jupiter's atmosphere (Figures~\ref{Fig:Maps} and~\ref{Fig:Pattern Maps}). Here we discuss the most important ones: 

(1) The Great Red Spot. The GRS is a stationary vortex embedded in the boundary between the SEB and the South Tropical Zone (STrZ). It was located at latitude $\rm 22\arcdeg$S and longitude about $\rm 245\arcdeg$W in early 2016 (Figure~\ref{Fig:DiskList}), but the location of the GRS location is not fixed on Jupiter. It longitudinally drifts along the southern interface of the SEB with a speed of $\rm \sim$2 $\rm m \cdot s^{-1}$, which is about $\rm \sim$120$\arcdeg$ per year \citep{simon2002new, simon2018historical}. The GRS is a region with enriched $\rm NH_{3}$ clouds and a high cloud top. Rising motions within the GRS are thought to bring gases ($\rm NH_{3}$, $\rm PH_{3}$, $\rm H_{2}$ with a low para-fraction, e.g., \citealt{fletcher2010thermal}) up from the deeper troposphere, and adiabatic expansion causes the GRS to be cold at the cloud-tops. It was proposed that at the cloud top $\rm NH_{3}$ photochemically reacts with the hydrocarbons (e.g., $\rm CH_{4}$, $\rm C_{2}H_{2}$) to produce chromophores that absorb UV and short-wavelength visible radiation (e.g., \citealt{ferris1987hcn, carlson2016chromophores}). Thus, on visible-wavelength maps, the GRS appears as a darker spot at shorter wavelengths (e.g., blue-absorbing chromophores) and brighter at longer wavelengths (high-altitude hazes, see reflection maps in Figures~\ref{Fig:Maps} and~\ref{Fig:Pattern Maps}). The additional gaseous and aerosol absorbers from the upwelling, combined with the cold upper tropospheric temperatures, make the GRS appear dark and cold at thermal-infrared wavelengths.    

(2) The North Equatorial Belt (NEB) expansion event. The NEB spans $\rm 7-17\arcdeg$N planetographic latitude and appears dark in reflected sunlight and bright (i.e., warm and cloud-free) in thermal-emission maps. The NEB periodically expands and contracts with a 3-5-year lifespan, and our early-2016 observations captured once such event \citep{fletcher2017jupiter}. As the expansion event occurs, the NEB expands northward into the North Tropical Zone (NTrZ), but the physical mechanism behind the NEB expansion event is not well understood---it may result from subsidence and cloud-clearing in the NTrZ associated with wave patterns on the northern edge of the NEB.

(3) Patchy cloud patterns in the North Temperate Belt (NTB). Patchy clouds in the NTB ($\rm 24-31\arcdeg$N) can be seen on the maps from 2015 to 2017 (Figures~\ref{Fig:Maps} and~\ref{Fig:Pattern Maps}). In February 2016, the dark-colored materials were distributed along the longitude ranging from $\rm \sim$170$\arcdeg$W to $\rm \sim$280$\arcdeg$W (reflection maps in Figures~\ref{Fig:Maps} and~\ref{Fig:Pattern Maps}). The white cloud patterns and brown chromophores interweave with each other in the NTB (visible wavelength maps in Figures~\ref{Fig:Maps} and~\ref{Fig:Pattern Maps}). The physical mechanism behind the intermittent NTB cloud is not well understood but is likely related to the regular plume outbreaks on the southern edge of the NTB (\citealt{sanchez2008depth, sanchez2017planetary}), the most recent of which occurred in October 2016.

(4) The South Equatorial Belt (SEB) outbreak. Although no major fade and revival cycles of the SEB have been detected since 2009-2011, smaller outbreaks of plumes do sometimes occur in the mid-SEB. One such outbreak occurred in Apr. 2017 at the longitude of $\rm \sim$250$\arcdeg$W (Figures~\ref{Fig:Maps} and~\ref{Fig:Pattern Maps}). The SEB outbreak was bright in the visible wavelengths, caused by the strong reflection of the enhanced local fresh clouds, but appears dark in the thermal infrared due to excess aerosol opacity and low temperatures associated with adiabatic expansion and cooling at the plume tops \citep{fletcher2017moist}. 

(5) Thermal waves and hot spots in the NEB. Jupiter's mid-NEB exhibits longitudinal thermal wave patterns with wavenumbers of 10-17 (\citealt{orton1991thermal, deming1997observations, li2006waves, fletcher2017jupiter}), where an undulating pattern of warm and cool spots is anti-correlated with dark and bright patches of reflected sunlight. This has been interpreted as aerosols condensing in the cooler upwelling regions of a Rossby wave and sublimating in the warmer subsiding branches. This wave pattern can be seen to be modulating the mid-NEB thermal emission at 10.3-10.5 $\rm \mu m$ in Figure~\ref{Fig:DiskList}.  In addition, the 5-$\rm \mu m$ hot spots are bright regions in thermal-emission wavelengths at the latitude $\rm 7\arcdeg$ to $\rm 10\arcdeg$N---the interface between the NEB and the Equatorial Zone (EZ). The 5-$\rm \mu m$ hot spots are a chain of cloud-free, cyclonic features controlled by a westward-propagating Rossby wave on the boundary between the EZ and the NEB (\citealt{showman2000nonlinear}). Although the 5-$\rm \mu m$ hot spots are mainly controlled by large-wavenumber Rossby waves, they are not equally bright---some hot spots are significantly brighter than the others. The brightness temperatures of warm regions (i.e., 5-$\rm \mu m$ hot spots) are about 240 to 260 K, meaning that the 5-$\rm \mu m$ observation probes the effective pressure level at $\rm \sim$3 bars (Section 5.2.3 in \citealt{west2004jovian} and Figure 3 in \citealt{fletcher2016mid}). Hot spots undergo different dynamical processes from that in the NEB. Numerical simulations suggest that there are strong vertical downdrafts in the hot spots, which could extend to at least a few bars. The downdrafts deplete the volatiles in this region \citep{showman2000nonlinear}. Thus, 5-$\rm \mu m$ hot spots are considerably brighter than the other regions on 5-$\rm \mu m$ maps (Figures~\ref{Fig:Maps} and~\ref{Fig:Pattern Maps}). Interestingly, Figure 4 in \cite{fletcher2016mid} shows the overlaps between several 5-$\rm \mu m$ hot spots and warm airmasses at 8.80 $\rm \mu m$ and 10.50 $\rm \mu m$. It is possible that the wave patterns (i.e., hot spots, warm airmsses) are controlled by the same group of Rossby waves.

(6) Small-scale patterns in belts and zones. There are numerous small-scale discrete patterns, such as anticyclonic white ovals near latitude $\rm 40\arcdeg$S, small cyclonic and anti-cyclonic patterns in the NEB and SEB. They ubiquitously exist in Jupiter's belts and zones. Although each individual small-scale pattern is too small to generate a large rotational variability. Their total contribution to the rotational modulation cannot be simply neglected. According to the latitudinal contribution profiles (left panels in Figures~\ref{Fig:Maps} and~\ref{Fig:Pattern Maps}), we crudely estimate that the contribution from the small-scale patterns is about 20\% to 30\%. However, because it is difficult to identify each individual pattern on the maps, we did not include them in the light-curve analysis in this study. Instead, the rotational modulations generated by these small-scale patterns are treated as the background variations.

Other than the horizontal structures discussed above, the vertical structure of Jupiter's atmosphere is another essential parameter to understand the rotational light curves at multiple wavelengths, because observations at different wavelengths could probe different altitudes. As we shall see later, the vertical structures of temperature, clouds (mainly $\rm NH_{3}$ clouds, possibly $\rm H_{2}O$ clouds and $\rm NH_{4}SH$ clouds), hazes and gases all play important roles in generating the rotational light curves of Jupiter. But their relative significances are wavelength-dependent.

\section{Jupiter's Light Curves}
\label{sec:Jupiter Light Curve}

Wavelength-dependent photometric variabilities are broadly exhibited by Jupiter's atmosphere from the UV to the mid-IR (Figure~\ref{Fig:Maps}). The light curves also show various kinds of non-sinusoidal shapes at different wavelengths (Figures~\ref{Fig:Maps},~\ref{Fig:Reflection Curves}, and~\ref{Fig:Emission Curves}). Figure~\ref{Fig:Reflection Curves} also shows a long-term light-curve evolution at reflection wavelengths over three years. Figure~\ref{Fig:Emission Curves} shows a short-term light-curve evolution at emission wavelengths over just a few weeks. 

Based on Jupiter's maps, we can interpret the shapes, amplitudes and evolution of Jupiter's light curves using the locations, sizes, brightness and temporal evolution of discrete patterns in belts and belt-zone interfaces, such as the GRS and the NEB expansion regions. Our analysis quantitatively shows that light curves are controlled by the size, local temperature contrast, cloud opacity with distributions of tropospheric gases, hazes and chromophores. The light-curve evolution is likely to be caused by the temporal evolution of periodic events such as the NEB expansion events and drifting of the stationary vortices such as the GRS and ovals.

\subsection{Light-Curve Amplitude}
\label{sec:light curve amplitude}

The light-curve amplitudes of Jupiter range from $\rm \sim$1-to-4\% from the UV to the mid-IR wavelengths, except for the light-curve amplitude at  5 $\rm \mu m$ which exceeds $\rm \sim$20\%, an order of magnitude larger than the others. This result is consistent with previous studies (\citealt{gelino2000variability, karalidi2015aeolus}). Based on the multi-wavelength maps, latitudinal contribution profiles and corresponding light curves, we can estimate the light-curve amplitudes by the sizes, brightness contrast ratios, and latitudes of the dominant patterns as:

\begin{equation} \label{eq6}
A \sim S_{p} \cdot \Delta B \cdot \cos{\theta},
\end{equation}
where $A$ is the light-curve amplitude; $S_{p}$ is the area fraction of the pattern over the disk; $\Delta B$ is the brightness contrast between the pattern and the background, which is given by $ \lvert B_{p}/{B_{b}}-1 \rvert$, where $B_{p}$ and $B_{b}$ are the brightness of the pattern and the background, respectively; $\cos\theta$ is the projection factor of the pattern on the disk where $\theta$ is the latitude of the discrete pattern. 

The size of the individual discrete pattern is almost independent of wavelength. For example, the north-south width and the east-west length of the GRS almost do not change at different wavelengths (Figure~\ref{Fig:Maps}). However, the brightness contrast is highly wavelength-dependent. The mechanism behind the brightness contrast at thermal-emission is different from that at reflection wavelengths. The reflection brightness contrasts between the discrete patterns and the background are controlled by the distributions of hazes, thin-thick clouds and chromophores in and above the clouds. The brightness contrasts at thermal-emission wavelengths are mainly produced by the distributions of temperature, opacities of the tropospheric gases, hazes and clouds. 

The brightness contrasts of the patterns vary from 20\% to 50\% from the UV to the mid-IR, except for that at 5 $\rm \mu m$, which can vary by a factor of 10 (5-$\rm \mu m$ hot spots, see color bars in Figure~\ref{Fig:Pattern Maps}). The light-curve amplitude at 5 $\rm \mu m$ is therefore significantly larger than any other wavelengths in this study. In the following sections, we first estimate the sizes of the discrete patterns. Then, we will focus on the mechanisms behind the wavelength-dependent brightness contrast.

\subsubsection{Size of the Weather Patterns}
\label{sec:pattern size}

\begin{figure}
\centering \includegraphics[scale = 0.75]{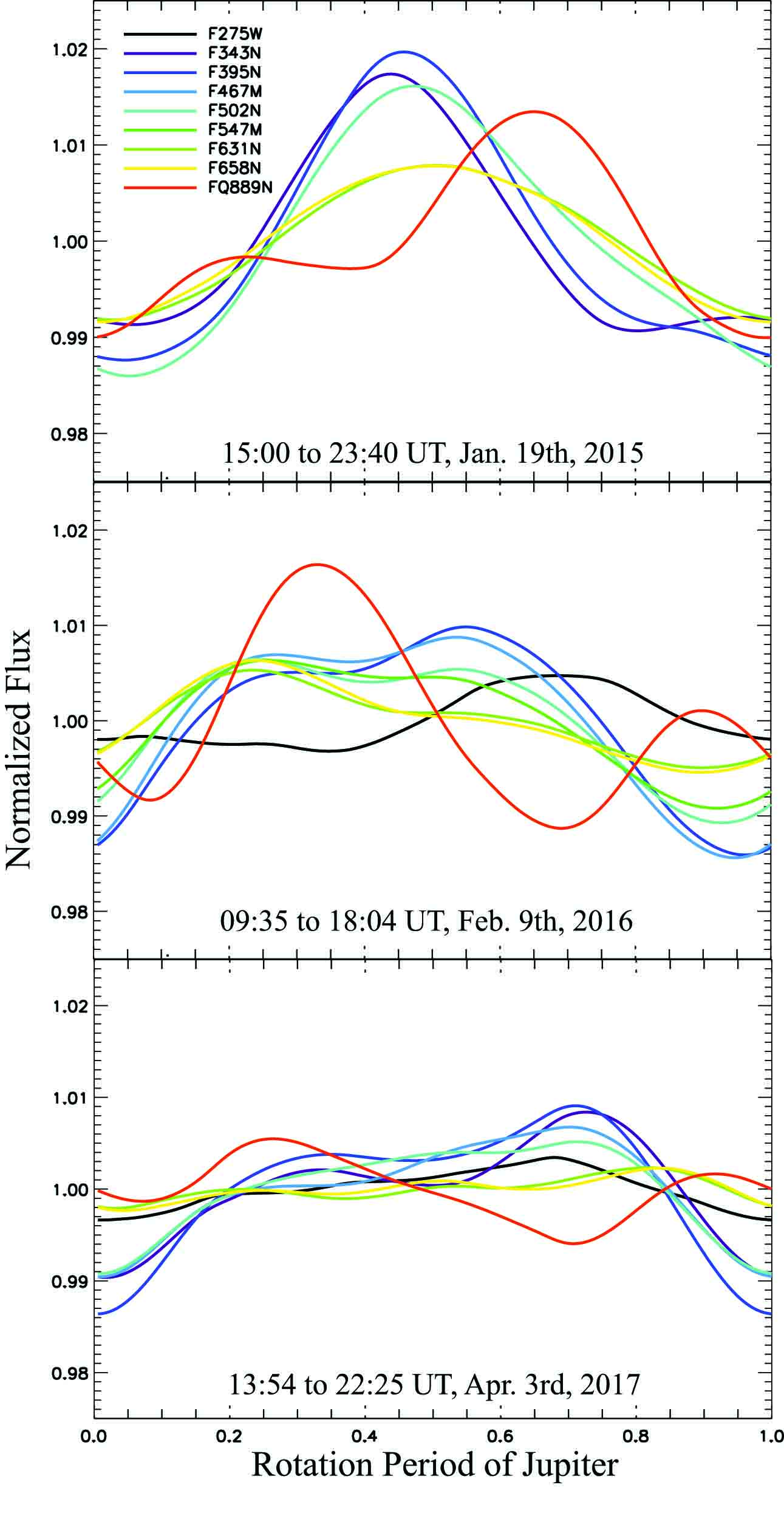} \
\caption{\label{Fig:Reflection Curves}Light curves of Jupiter from 2015 to 2017 at reflection wavelengths. Most simultaneously observed light curves have similar shapes. The methane-band light curves (FQ889N) have obvious opposite shape behavior compared with light curves at the other wavelengths. The amplitudes of the long-wavelength (e.g., F658N) visible light curves are smaller than the short-wavelength ones (e.g. F343N).}
\end{figure}

The light-curve contribution maps and latitudinal contribution profiles (left panels in Figures~\ref{Fig:Contribution Maps} and~\ref{Fig:Pattern Maps}) indicate that, in February 2016, the important patterns are mainly located in the SEB, NTB, and the northern and southern boundaries of the NEB. The corresponding discrete patterns are the GRS, patchy clouds in the NTB, the NEB expanded region (which does not cover all longitudes), and 5-$\rm \mu m$ hot spots controlled by planetary waves, respectively. Here, we focus on the most important two features: NEB expansion event and the GRS.

When the NEB expansion occurred in 2016, the expanded region became one of the most important discrete patterns at both reflection and thermal-emission wavelengths (Figures~\ref{Fig:Contribution Maps} and~\ref{Fig:Maps}). During the expansion event, the dark-colored materials in the NEB appeared to expand northward (see reflection maps in Figures~\ref{Fig:Maps} and~\ref{Fig:Pattern Maps}) into the NTrZ \citep{fletcher2017jupiter}. Usually, the expansion occurs uniformly at all longitudes. But in 2016, only about one third of the full circumference of the NEB is expanded northward ($\rm \sim$50$\arcdeg$W to $\rm \sim$290$\arcdeg$W in the system III longitude) \citep{fletcher2017jupiter}. This expansion region has a north-south width of $\rm \sim$3$\arcdeg$, west-east length of $\rm \sim$120$\arcdeg$W. The size, $\rm S_p$, of the NEB expansion region is about 3\%. 

The GRS is important for the rotational modulation from the UV to the mid-IR on Jupiter. This is consistent with the results in \cite{karalidi2015aeolus}. The size of the GRS core is well measured. The north-south width and east-west length are about 17,000 km and 39,000 km, respectively \citep{ingersoll2004dynamics}. The size and the brightness contrast of the whole GRS are not well constrained, because the GRS is not a simple dark spot on Jupiter maps (Figures~\ref{Fig:Maps} and~\ref{Fig:Pattern Maps}). There are $\rm NH_{3}$ cloud spots and turbulent activities toward the northwest of the GRS, a region known as the ``turbulent wake''. This cloudy region is associated with the dynamic processes of the GRS \citep{palotai20143d}. Based on the bright GRS on the methane-band map (Figures~\ref{Fig:Maps} and~\ref{Fig:Pattern Maps}), we estimate that the area fraction of the GRS, $\rm S_p$, is 5\%.

\subsubsection{Brightness Contrast at Reflection Wavelengths: Aerosols, Chromophores and Methane}
\label{sec:reflectioncontrast}

The brightness contrasts of discrete patterns on reflection maps vary from 20\% to 50\% (see colorbars in Figure~\ref{Fig:Maps}). The reflection brightness is dominated by the reflection of $\rm NH_{3}$ clouds and the absorption of hazes, chromophores in the $\rm NH_{3}$ clouds, and the methane gas absorption above the clouds. 

The haze absorption dominates the appearance and the brightness contrast of Jupiter in the UV (275 nm). The GRS is darker than the background due to the particles on the cloud top of the GRS \citep{west2004jovian,zhang2013stratospheric, carlson2016chromophores}. At other latitudes, the distributions and brightness of upper tropospheric hazes are nearly uniform in longitude (UV map in Figure~\ref{Fig:Maps}). Furthermore, the brightness contrast between belts and zones is less than 20\%, which is significantly smaller than that at other wavelengths (Figure~\ref{Fig:Maps}). But the NEB expansion event could still be vaguely seen from longitude $\rm \sim$80$\rm \arcdeg$W to $\rm \sim$180$\rm \arcdeg$W. Given that the dominant pattern covers about a few percent, the amplitude of the UV light curve is therefore less than $\rm \sim$1\% (Equation~\ref{eq6}).

The brightness contrast in the visible maps is dominated by chromophores in and above the $\rm NH_{3}$ clouds. Recent studies indicate that the reddish-brown chromophores ubiquitously exist in Jupiter's upper troposphere \citep{carlson2016chromophores, sromovsky2017possibly}. The chromophores preferentially absorb short-wavelength visible light, and the absorbability generally decreases as the wavelength increases. The column density of chromophores in the NEB, SEB and GRS is about 18 to 20 $\rm \mu g\cdot cm^{-2}$, while the column density in the EZ is about 13 $\rm \mu g\cdot cm^{-2}$ \citep{sromovsky2017possibly}. Because of abundant chromophores, belts and the GRS absorb more short-wavelength visible light than the zones. The brightness contrasts between belts and zones decrease from $\rm \sim$50\% at 395 nm to $\rm \sim$20\% at 658 nm. Accordingly, because the pattern size is independent of wavelength, light-curve amplitude decreases as wavelength increases (Figures~\ref{Fig:Maps} and~\ref{Fig:Reflection Curves}). 

The methane-band map is controlled by methane absorption as well as $\rm NH_{3}$ cloud reflection. The effective pressure level at 889 nm is $\rm \sim$0.6 bars \citep{west2004jovian}, corresponding to the $\rm NH_{3}$ cloud top in the EZ and the GRS. Aerosols above this level lead to more cloud reflection and less $\rm CH_{4}$ absorption. Thus, the EZ and the GRS appear brighter than the NEB and SEB (see methane-band map in Figure~\ref{Fig:Maps}). Because methane is well-mixed, the spatial brightness contrast is mainly controlled by the spatial distribution of the $\rm NH_{3}$ clouds. The methane-band map shows that the brightness contrast between the GRS and the SEB is roughly 50\%, leading to a light-curve amplitude of $\rm \sim$2\%. Furthermore, because belts and the GRS appear brighter than the surroundings at 889 nm instead of darker in the visible wavelengths, the shape of the 889-nm light curve is significantly different from the others. We will discuss the light-curve opposite-shape behavior in methane-band light curve in Section~\ref{sec:shape}.

\subsubsection{Brightness Contrast at Emission: Temperature vs Opacity}
\label{sec:emission contrast}

At 8.80 $\rm \mu m$ and 10.50 $\rm \mu m$, the radiance contrast is roughly 50\% (Figures~\ref{Fig:Contribution Maps} and~\ref{Fig:Maps}), implying that the brightness temperature contrast is about 6 K \citep{fletcher2016mid}. However, at 5 $\rm \mu m$, the radiance contrast could exceed by a factor of 5 (Figure~\ref{Fig:Maps}). The corresponding brightness temperature contrast is larger than 20 K \citep{fletcher2016mid}. 

The brightness temperature contrast is primarily controlled by the spatial variations of temperature, gases, and aerosols at 8.80 $\rm \mu m$ and 10.50 $\rm \mu m$. The former wavelength is more sensitive to upper tropospheric hazes and phosphine, whereas the latter one is more sensitive to the tropospheric $\rm NH_{3}$ gas. As an example, here we quantitatively estimate the brightness temperature contrast induced by the horizontal distributions of $\rm NH_{3}$ gas and temperature at 10.50 $\rm \mu m$. At the same pressure level, $\rm NH_{3}$ is not homogeneously distributed over the globe. The retrieved $\rm NH_{3}$ mixing ratio at $\rm \sim$500 mbars in the GRS and the EZ is roughly 20 to 25 ppm, about twice of that in the SEB and NEB \citep{fletcher2016mid}. Therefore, the effective emission levels (where the gas optical depth is unity) in the $\rm NH_{3}$ enriched region (e.g., the GRS/NTrZ) is higher than that in the $\rm NH_{3}$ depleted region (e.g., the NEB/SEB). Observations show that $\rm NH_{3}$ mixing ratio exponentially decreases from 100 ppm at 700 mbars to 0.1 ppm at 250 mbars \citep{showman2005dynamical, fletcher2016mid}. Using this profile, we estimate that the height difference is about 2 km between the effective emission levels in the $\rm NH_{3}$ enriched and depleted regions. Given the adiabatic lapse rate of $\rm \sim$2 $\rm K\cdot km^{-1}$, the temperature difference between the two effective emission levels (due to the gas opacity contrast) is about 3 to 4 K. On the other hand, the horizontal temperature difference between the NEB and the NTrZ, as well as that between the GRS and the SEB, is also $\rm \sim$3 K \citep{nixon2010abundances, fletcher2016mid} at the this pressure level ($\rm \sim$500 mbars). Therefore, the total brightness temperature contrast between the GRS/SEB and the NEB/NTrZ, induced by both gas opacity and horizontal temperature contrast, is about 5 to 7 K. This leads to about 40\% to 60\% brightness contrast at 10.50 $\rm \mu m$ (see colorbar in Figure~\ref{Fig:Maps}). Given that the dominant pattern size, $\rm S_p$, is about 3\% to 5\%, the magnitude of the rotational modulations at 10.50 $\rm \mu m$ is therefore $\rm \sim$2 to 4\%, in agreement with the observed rotational modulation at 10.50 $\rm \mu m$ (Figures~\ref{Fig:Maps} and~\ref{Fig:Reflection Curves}).

Observations at 5 $\rm \mu m$ are considerably different from other thermal-emission wavelengths. 5-$\rm \mu m$ maps are mainly dominated by brightness contrast induced by vertical cloud structures. The latitudinal-contribution profiles indicate that the 5-$\rm \mu m$ hot spots provide a significant contribution to the light-curve amplitude. Because at 5-$\rm \mu m$ light can penetrate into deep troposphere in the cloudless hot-spot region but cannot go through the thick $\rm NH_{3}$ cloud layers in zones, the observations roughly probe three different layers. The highest layer is the cloudy zones at 700 mbars with thick $\rm NH_{3}$ cloud region. The middle layer is the belts with thin clouds deeper than that in the thick cloud region (i.e., the GRS, zones). The deepest layer is the deep atmospheric layer seen in the 5-micron hot spot at about 1-4 bars. Thus, the 5-$\rm \mu m$ hot spots can be considered as $\rm NH_3$ cloud holes. The height difference between the effective pressure level of 5-$\rm \mu m$ hot spots and the effective pressure level of the background (i.e., interface between the NEB and the EZ) is larger than $\rm \sim$10 km. The corresponding temperature difference between two layers is larger than 20 K, which is sufficient to produce a large brightness contrast and a resultant large rotational modulation. Although there is no good observational constraint on the temperature distribution below the $\rm NH_{3}$ cloud layers, simulations suggest that the largest horizontal temperature contrast at the same pressure level is limited below 5 K \citep{lian2010generation}. Therefore we conclude that the large light-curve amplitude at 5 $\rm \mu m$ is generated from the 5-$\rm \mu m$ hot spots (i.e., the $\rm NH_{3}$ cloud holes) and thin-thick cloud distributions instead of the horizontal temperature variation at the same pressure level.

\subsection{Light Curve Shape, Evolution and Phase Shift}

\subsubsection{Light Curve Shape}
\label{sec:shape}

\begin{figure}
\centering \includegraphics[scale = 0.75]{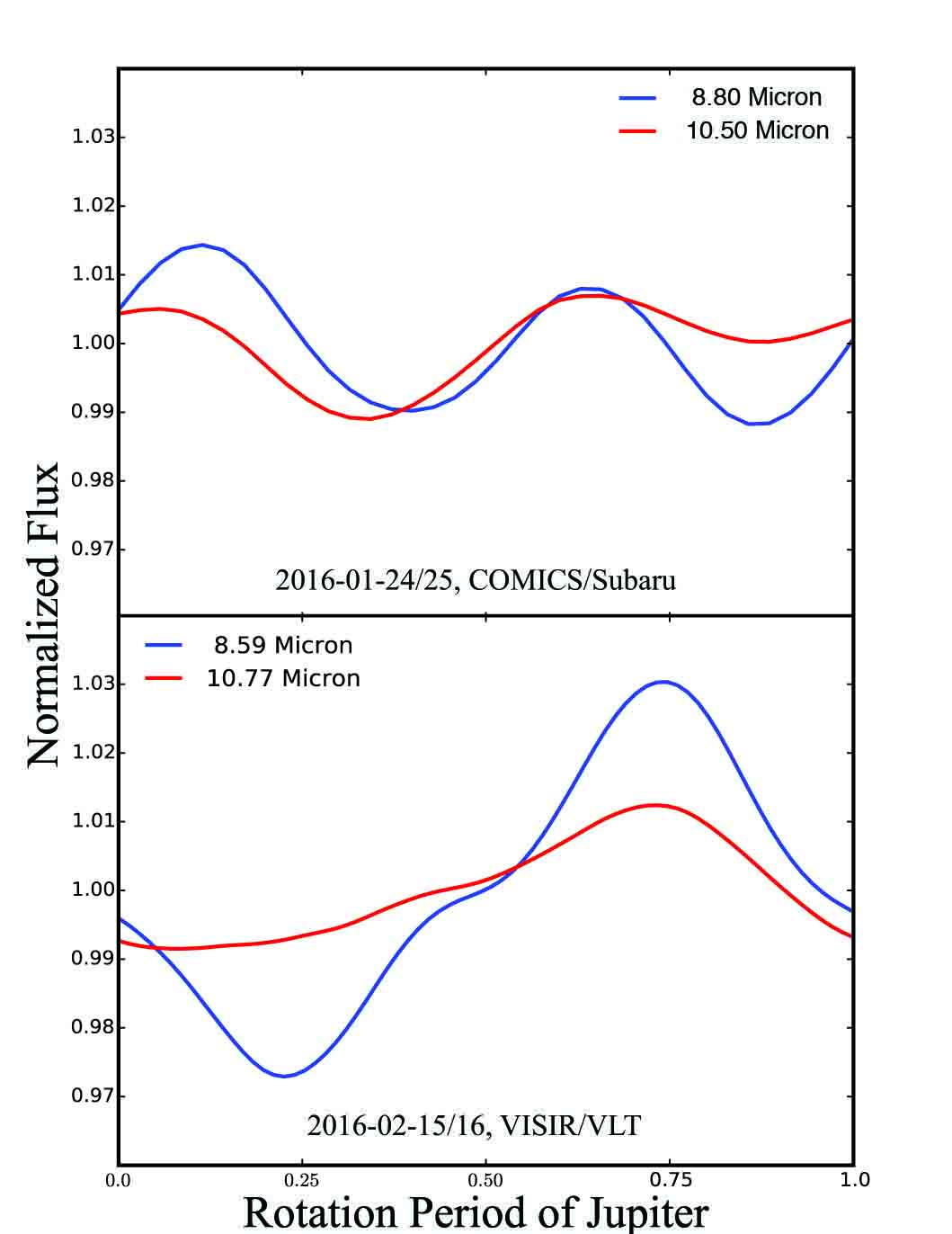} 
\caption{\label{Fig:Emission Curves}Jupiter's thermal-emission light curves separated by three weeks. The light-curve amplitude increases by a factor of 2 in a timescale of $\rm \sim$50 Jupiter's rotations. The locations of the peaks and troughs of the two light curves from the same observational date are similar. But their shapes evolve from a two-peak structure to a single-peak structure just after three weeks.}
\end{figure}

Most simultaneously imaged light curves have similar shapes, but are often irregular instead of sinusoidal (Figures~\ref{Fig:Reflection Curves} and~\ref{Fig:Emission Curves}). At reflection wavelengths, light curves in January 2015 show one obvious peak near the longitude of $\rm \sim$180$\arcdeg$W. But, in February 2016 and April 2017, except for the methane band (889 nm), light curves show an obvious trough instead of a peak near the longitude of $\rm \sim$30$\arcdeg$W. The methane-band light curve in February 2016 has 2 peaks near longitude $\rm \sim$240$\rm \arcdeg$W and $\rm \sim$30$\rm \arcdeg$W, and 2 troughs at longitude $\rm \sim$330$\rm \arcdeg$W and $\rm \sim$110$\rm \arcdeg$W. In April 2017, it also has a two-peak structure. 

The light-curve peaks and troughs are not precisely located at the same longitudes of the discrete patterns. For example, in February 2016, at the visible wavelengths (Figure~\ref{Fig:EvolutionMaps}), the small light curve trough at the longitude $\rm \sim$215$\rm \arcdeg$W does not coincide with the location of the GRS at $\rm \sim$245$\rm \arcdeg$W. In April 2017 (Figure~\ref{Fig:EvolutionMaps}), the trough of the visible light curve appears at the longitude about $\rm \sim$0$\rm \arcdeg$W, but the longitude of the GRS is at $\rm \sim$25$\rm \arcdeg$W. The reason is that none of the discrete patterns is simply dark or bright compared to the background (as noted in Section~\ref{sec:pattern size}). For example, the bright clouds on the west limb of the GRS increase the total reflected flux and shift the light-curve trough eastward. The numerous small-scale structures in the belts and zones (Section~\ref{sec:Jupiter maps}, point 6), which are not well resolved in our current images, can also modulate light curves. In addition, the contribution from dark patterns and bright patterns can cancel out each other. According to the light-curve contribution maps at 395 nm (Figure~\ref{Fig:Contribution Maps}), the contribution from the GRS offsets that from the bright regions in the west limb of the GRS and the NEB expand area (at the rotation period from 0.2 to 0.4). Thus, the locations of important discrete patterns cannot be precisely identified from the light-curve shapes.

\subsubsection{Light Curve Evolution}
\label{sec:evolution}

\begin{figure*}
\centering \includegraphics[width = 1.0\textwidth]{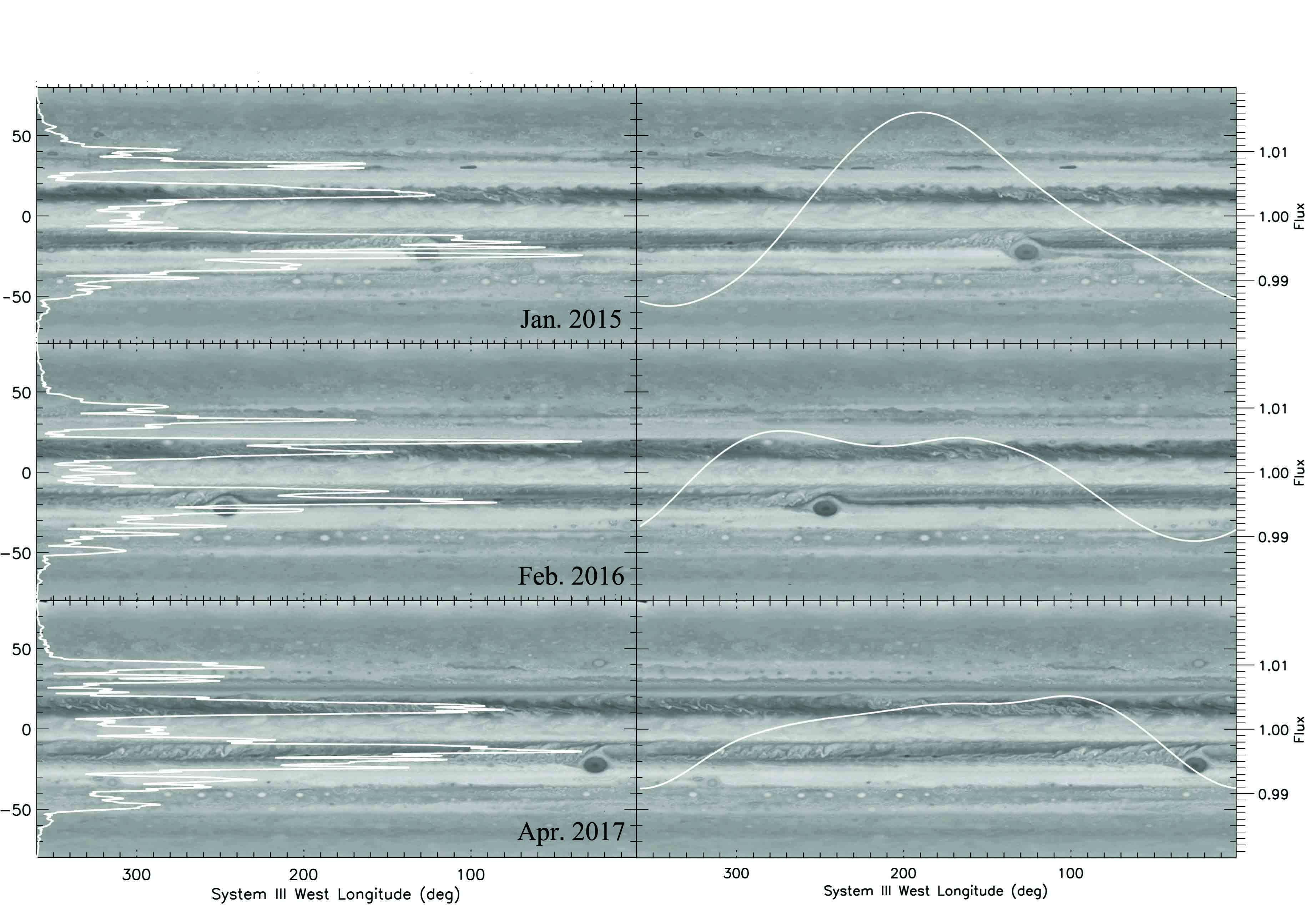} 
\caption{\label{Fig:EvolutionMaps}Long-term evolution of Jupiter's visible maps and rotational light curves (white lines in the right panels) at 395 nm from 2015 to 2017. The latitudinal contribution profiles (white lines in the left panels) indicate that the contribution to the rotational modulation at each latitude does not change much with time. The maps also show the evolution of several periodic events (e.g., NEB expansion event, SEB outbreak) and drifting of the GRS, as well as the drifting of the patchy clouds in the NTB. The occurrence and disappearance of the NEB expansion event, which do not progress equally over all longitudes, can also be seen from the latitudinal contribution profiles.}
\end{figure*}

Jupiter's light curves exhibit active short-term and long-term evolution (Figures~\ref{Fig:Reflection Curves},~\ref{Fig:Emission Curves}, and~\ref{Fig:EvolutionMaps}). The short-term evolution is seen during $\rm \sim$50 Jupiter rotations between January and February of 2016 at thermal-emission wavelengths (Figure~\ref{Fig:Emission Curves}). The long-term evolution is seen at reflection wavelengths from 2015 to 2017 (Figures~\ref{Fig:Reflection Curves} and~\ref{Fig:EvolutionMaps}). The light-curve evolution is a result of the evolution of periodic events, such as the NEB expansion and SEB outbreak, the longitudinal drifting of the GRS, and the patchy clouds in NTB. However, we do not have continuous long-term observations to unveil sufficient details about the short-term and long-term light-curve evolution on Jupiter. More observations are needed in the future to quantify the underlying mechanisms.

\subsubsection{Wavelength Dependence}
\label{sec:phase shift}

As noted in Section~\ref{sec:shape}, although most simultaneously imaged light curves exhibit similar shapes, the light curve at the methane-band wavelength (889 nm) is significantly different from that in the other wavelengths (Figures~\ref{Fig:Maps} and~\ref{Fig:Reflection Curves}). For example, in January 2015, February 2016, and April 2017, the methane-band light curves exhibit peaks where the light curves in the UV and visible wavelengths have troughs (Figure~\ref{Fig:Reflection Curves}).

The opposite shape behavior of the methane-band light curve is a result of the brightness change of the GRS at different wavelengths. Normally the GRS appears dark at visible wavelengths due to chromophore absorption. However, because the methane absorption is very strong at 889 nm, the GRS appears bright because there is less methane absorption above the high cloud top of the GRS. Therefore, when the GRS rotates into the view, the total flux decreases at visible wavelengths but increases at 889 nm. As a result, the methane-band light curves exhibit an opposite-behavior to the other wavelengths.

Light curves at different visible wavelengths also appear slightly differently (Figure~\ref{Fig:Reflection Curves}). These behaviors are primarily induced by the chromophores. Because chromophores absorb more short-wavelength visible light (that is why they appear reddish), they create larger brightness contrast between the chromophore-enriched patterns and background at shorter wavelengths. 

The amplitudes of the two simultaneously observed light curves at thermal emission (8.80 $\rm \mu m$ and 10.50 $\rm \mu m$, see Figure~\ref{Fig:Emission Curves}) are different by a factor of 2, but the locations of peaks and troughs coincide with each other. A possible reason is that these two wavelengths probe roughly the same pressure level with similar distributions of temperature, aerosols and gases ($\rm NH_{3}$ and $\rm PH_{3}$). The shape of the 5-$\rm \mu m$ light curve is also different from the other wavelengths. This might be expected because the underlying mechanism (i.e., cloud holes, thin-thick cloud structures) of the 5-$\rm \mu m$ light curve is very different from the other wavelengths. But note that our 5-$\rm \mu m$ images are taken at a different date, so the light-curve differences might also be a result of time evolution. More observations are needed in the future to investigate these behaviors in detail.

\section{Implications for Brown Dwarfs and Exoplanets}
\label{sec:implications}

It is believed that temperature perturbations and patchy cloud structures are responsible for the complex photometric rotational variabilities observed on brown dwarfs and directly imaged exoplanets (e.g., \citealt{marley2010patchy, buenzli2012vertical, apai2013hst, morley2014spectral, esplin2016photometric, yang2016extrasolar, zhou2016discovery, apai2017zones}). Weather patterns on brown dwarfs have been inferred by some indirect methods, such as mapping tools and Doppler imaging (\citealt{crossfield2014global}; \citealt{karalidi2015aeolus}; \citealt{apai2017zones}). However, without directly resolved images, it is difficult to understand the real mechanism underlying the rotational variability and light-curve evolution. This study of Jupiter's light curves could provide important insights on how the rotational modulations are produced in cloudy atmospheres. Here, we list some implications on brown dwarfs and exoplanets from the perspectives of planetary waves, pattern size, the significances of temperature and opacity sources, and the wavelength-dependent behavior.

\subsection{Patterns Controlled by Planetary-scale Waves}
\label{sec:implication of wave}

\cite{apai2017zones} suggested that the beating of planetary-wave-controlled patterns on brown dwarfs is a dominant cause for the rotational modulations and light-curve evolution. On Jupiter, the large-wavenumber (i.e., small-wavelength) patterns with unequally bright patterns, such as the bright bulges controlled by the tropospheric thermal waves (\citealt{ingersoll2004dynamics}; \citealt{fletcher2017jupiter}) and 5-$\rm \mu m$ hot spots controlled by Rossby waves, can generate rotational modulations at 5 $\rm \mu m$, 8.80 $\rm \mu m$ and 10.50 $\rm \mu m$, respectively (Figure~\ref{Fig:Maps}). At 5 $\rm \mu m$, hot spots are the dominant patterns for the rotational variability. But note that high-wavenumber planetary waves and associated weather patterns such as hot spots, if they have equal brightness, are not likely to induce a large rotational modulation because the brightness peaks and troughs tend to cancel out each other when averaged over the disk. Weather patterns with large-scale longitudinal asymmetry as seen around the NEB on Jupiter, which might be induced by low-wavenumber planetary waves or other complex meteorological activities, could also be responsible for the large rotational modulations on brown dwarfs and exoplanets. 

\subsection{Pattern Size}
\label{sec:implication of size}

The GRS and the NEB expansion region (provided it does not progress to be longitudinally uniform) are found to be the important discrete patterns contributing to the rotational modulations of Jupiter. Our study shows that pattern size is very important in determining the magnitude of the photometric variability. Empirically, the pattern size on Jupiter is strongly related to the width of belts and zones. As shown in Figures~\ref{Fig:Maps} and~\ref{Fig:Pattern Maps}, the north-south width and the west-east length of the GRS are comparable to the width of the belt and zones. Because the NEB expansion event occurs between two adjacent belts and zones (the NEB and the NTrZ), the width of expansion region should be smaller than the width of the bands. In an atmospheric mapping model, Aeolus, \cite{karalidi2015aeolus} assumed the diameter of the spot is equivalent to the width of the band. \cite{zhang2014atmospheric} suggested that the atmospheres of cold brown dwarfs could be dominated by the banded structure. If this were true, the band width might be estimated by the Rhines scale $\rm L_{\beta} \sim \pi(U/\beta)^{\frac{1}{2}}$, where $\beta$ is the Rossby parameter $\rm2\,\Omega\cos\phi/a$; $\phi$ is the latitude; a is planet radius; U is the characteristic flow speed (\citealt{rhines1975waves}; \citealt{vasavada2005jovian}). Therefore, the size of a large pattern might be empirically given by $\rm L_{\beta}/a\,\sim\sqrt{\,{\pi}^2 U/2\,\Omega\,a\cos\phi}$. For a Jupiter-size brown dwarf ($\rm {\sim}10^{5}\,km$) with a typical spinning rate of several hours (\citealt{yang2016extrasolar}) and a flow speed of $\sim$100 $\rm m \cdot s^{-1}$ (see numerical simulation results in \citealt{showman2013atmospheric}, \citealt{zhang2014atmospheric} and \citealt{tan2016atmospheric}), we estimate that the size of a single discrete pattern might be smaller than 10\% of the disk area. This is consistent with our Jupiter study---the GRS area fraction is about 5\%.

\subsection{Temperature versus Opacities of Gas and Clouds}
\label{sec:implication of T vs Tau}

Previous studies only proposed that patchy cloud opacity and temperature perturbations as a source for the rotational modulation on brown dwarf atmospheres (e.g., \citealt{apai2013hst}; \citealt{zhang2014atmospheric}; \citealt{morley2014spectral}). On the other hand, our study on Jupiter shows that the distributions of gas, $\rm NH_{3}$ clouds, aerosols and horizontal temperature in belts and zones are all important for the rotational modulations in thermal emission. The brightness contrast is primarily produced by the $\rm NH_{3}$ cloud holes and thin-thick clouds at 5 $\rm \mu m$ (Section~\ref{sec:emission contrast}). The GRS, which is a localized high-cloud-top region, can generate rotational variability at all wavelengths. The GRS scenario is similar to the thin-thick cloud scenario proposed by \cite{apai2013hst}. But we found that the horizontal variations of gas opacity are also important for the thermal-emission light curves on Jupiter, a new mechanism that was not considered in previous brown dwarf studies. 

On Jupiter, the distributions of temperature, gas opacity and cloud opacity can significantly influence the light-curve amplitude. The rotational modulations at both 8.80 $\rm \mu m$ and 10.50 $\rm \mu m$ are primarily controlled by the distributions of temperature and gas opacity. The light-curve amplitudes of Jupiter at two thermal emission wavelengths are 2-4\%. We note that the light-curve amplitudes of most variable brown dwarfs are also on the order of a few percent, implying that a similar mechanism---temperature and gas variation---might be applied to these low-amplitude-variability brown dwarfs. On the other hand, at 5 $\rm \mu m$, the large light-curve amplitude ($\rm \sim$20\%) of the rotational modulation is instead dominated by the cloud opacity variation (i.e., cloud holes, thin-thick clouds) because the 5-$\rm \mu m$ observation is able to probe into the hot and deep atmosphere in the cloud-free region. Compared with the cases of 8.80 and 10.50 $\rm \mu m$, the horizontal temperature variation is negligible to the rotational modulation at 5 $\rm \mu m$. Some brown dwarfs (e.g., 2MASS2139, 2MASS1324) indeed show high-amplitude light curves at certain wavelengths (\citealt{radigan2012large, heinze2015weather}). We suggest that the underlying mechanism relevant to these brown dwarfs might be similar to that on Jupiter at 5 $\rm \mu m$. We also note that the light-curve amplitudes of some brown dwarfs behave differently at different wavelengths (e.g., 2MASS2139 in \citealt{radigan2012large, apai2013hst}). This might indicate that the relative significances of temperature, gas and cloud opacities are wavelength-dependent on these bodies, just like their bahaviors on Jupiter.

\cite{morley2014spectral} discussed the contribution of temperature perturbation and patchy clouds to the rotational modulation. But note that the distributions of temperature, gas, and clouds could be highly correlated when the gas is condensable, because the condensable gas tends to form more clouds in cold regions, and clouds could have radiative feedback to modulate the local temperature. Alternatively, the gas abundances, clouds, and temperature distributions could all be controlled by vertical motions in convective atmospheres. For example, Jupiter's belts are less cloudy and warm, while zones are cloudy and cool. But even on Jupiter, why the clouds are enriched in zones and depleted in belts, and how the cloud abundance is related to the banded structure are still not well understood.

\subsection{Light-Curve Wavelength Dependence}
\label{sec:phase shift implication}

It is reported that some brown dwarfs exhibit wavelength-dependent rotational variabilities. The observations suggest that the ``phases'' of the light curves at some wavelengths are shifted compared with other wavelengths (e.g., \citealt{buenzli2012vertical, yang2016extrasolar}). On Jupiter, we do see interesting wavelength-dependent behaviors, such as various light-curve shapes and the peak shift of the methane-band light curve. However, we do not see an obvious ``phase shift'', because the multi-wavelength light-curve shapes are highly wavelength-dependent and significantly different from each other. The reason is that, as discussed in \cite{karalidi2015aeolus}, Jupiter's light curves have a much higher signal-to-noise ratio compared with observed brown dwarf observations, showing more detailed structures in their shapes. With more observations on brown dwarfs in the future, we may see more details of their light curves. We suggest that the mechanism behind the currently observed light-curve ``phase shifts'' on brown dwarfs is similar to that controls the wavelength-dependent behaviors (i.e., light-curve peak shift, opposite-shape behaviors) on Jupiter. For example, the observed ``phase shifts'' can be induced by the wavelength-dependent brightness of some patterns such as the GRS on Jupiter.

\subsection{Inclination Angle Dependence}
\label{inclination}

The observed rotational modulation might also depend on the inclination angle to the observer. Here we define the inclination angle is zero for equator-on objects and inclination angle is 90$\rm \arcdeg$ for pole-on objects. We expect that on a low-inclined body, such as Jupiter to an observer on the Earth, the spatial variations in the equatorial and tropical regions that contribute the most to the disc-integrated flux, would be detectable because of rotational variations. For a planet observed pole-on, there is no variability in time that stems from rotational variations; only temporal modulations would be detected. Therefore, we would see the brightness modulations caused by different physical mechanisms. In fact, both the observations in \cite{vos2017viewing} and simulation results in \cite{kostov2012mapping} have shown that equator-on brown dwarfs have larger rotational modulations than the pole-on.

\section{Conclusions and Discussion}
\label{sec:conclusions}

In this study, we constructed and analyzed Jupiter's rotational light curves from the UV to the mid-IR wavelengths. The photometric variability and light-curve evolution are broadly exhibited at all wavelengths. Light-curve amplitudes and shapes are wavelength-dependent. In our analysis, the light-curve amplitudes vary from 1\% to 4\% at reflection wavelengths, 8.80 $\rm \mu m$, and 10.50 $\rm \mu m$, but the 5 $\rm \mu m$ light-curve amplitude is more than 20\%. We used the latitudinal contribution profiles to identify the locations of the discrete patterns. The results show that the rotational modulations are primarily produced in the belts. Several important discrete patterns have been identified: (1) GRS; (2) NEB expansion event in 2016 (which did not extend over all longitudes); (3) patchy cloud patterns in the NTB; (4) planetary-wave-controlled hot spots; (5) SEB outbreaks; (6) small-scale patterns in belts and zones.

Our results show that the amplitudes of light curves are controlled by the brightness contrasts, sizes and latitudes of the discrete patterns. The sizes of the dominant patterns (e.g., the GRS and the NEB expanded region) are about 3\% to 5\% of Jupiter's disk area. The discrete pattern size is almost independent of wavelength. 

We found that a large photometric variability tends to occur at a wavelength with a large brightness contrast on the map. At the reflection wavelengths, the spatial brightness contrast is controlled by tropospheric haze distribution, the absorption of chromophores and patchy $\rm NH_{3}$ clouds. In the UV, the brightness contrast is small due to the nearly uniform distributions of tropospheric hazes as a function of longitude. At the visible wavelengths, because chromophores absorb short-wavelength visible light and reflect the long-wavelength light, the brightness contrast is smaller at longer wavelengths, leading to a smaller rotational modulation. In the methane band (889 nm), because the methane absorption is smaller in the high-cloud-top regions, the distributions of the patchy clouds, such as the GRS, dominate the brightness contrast. 

In thermal emission, the brightness contrasts are generated by the inhomogeneous distributions of temperature, clouds, and gases. The relative significances of these factors are wavelength-dependent. At 8.80 $\rm \mu m$ and 10.50 $\rm \mu m$, both the temperature contrast and gas opacity are important because the observations probe the pressure level above the main cloud layers. At the 5-$\rm \mu m$ atmospheric window, observations can probe down to 1-to-4 bars in $\rm NH_{3}$ cloud holes and less than 1 bar in the thick-cloud regions. The cloud opacity could therefore play a controlling role in the brightness contrast and the large rotational modulation at 5 $\rm \mu m$.

We also found various light-curve shapes, short-term and long-term light-curve evolution on Jupiter. The shapes of the light curves are dominated by the locations and shapes of the patterns, such as the GRS, the NEB expanded region, and small-scale cyclonic patterns, in the NEB and the SEB. But the light-curve peaks and troughs do not precisely coincide with the locations of the discrete patterns. The light-curve evolution is controlled by the temporally varying dynamic patterns, such as the drifting of the GRS, the periodic NEB expansion event, and the SEB outbreak event.

The light curves at the 889-nm methane band show an opposite-shape behavior compared with the other wavelengths. The GRS appears as a bright spot in this methane band but as a dark spot at the UV and visible due to the high cloud top at the GRS. As a result, light curves in the methane band (889 nm) vary in the opposite sense to that at the other wavelengths.

Our study provides important insights to the study of the photometric variability on brown dwarfs and directly imaged exoplanets. We suggest the following. (1) If the brown dwarfs and the planets are dominated by banded structures, and if we assume the sizes of cloudy spots and band expansion are limited by the width of the bands, the size of discrete patterns is probably smaller than the Rhines scale. (2) The distributions of the temperature, gas opacity and patchy cloud opacity control the brightness contrast. For the wavelengths that sense the pressure levels above the main region of condensed volatiles, the distributions of temperature, gases, and aerosols should be important to the rotational modulation. This mechanism might be related to brown dwarfs with low-amplitude light curves. The cloud holes are more likely to produce large photometric variability at the wavelengths of atmospheric windows on brown dwarfs. We also pointed out that, because of the dynamical correlation (i.e., cooling or heating effects caused by the convections) between the cloud formation and temperature, there could be a degeneracy in the light-curve signal between the horizontal temperature contrast and patchy cloud opacity. (3) Bright and dark bulges with large-scale brightness structures could be the causes of the rotational modulations. These unequal brightness sources might be induced by large-scale planetary waves and other weather activities. (4) For brown dwarfs observed pole-on, we should see less rotational variations and more dynamical variations; for brown dwarfs observed equator-on, we should see more rotational variations and a resultant larger light-curve amplitude.

Finally, in the future, multi-wavelength (especially at the thermal wavelengths), continuous photometric monitoring of Jupiter for several successive rotations should provide more details on the underlying mechanisms of Jupiter's rotational light curves and their time evolution. It will further shed light on our interpretation of the rich observational data of rotational modulation on brown dwarfs and directly imaged exoplanets.

\section{Acknowledgments}
We thank Theodora Karalidi, Michael C. Liu and Ji Wang for helpful discussions. This research was supported by NASA Earth and Space Sciences Fellowship to H.G. and NASA Solar System Workings grant NNX16AG08G as well as the Hellman Fellowship to X.Z. This research is also benefited from the Outer Planetary Atmosphere Legacy project at https://archive.stsci.edu/prepds/opal/. LNF was supported by a Royal Society Research Fellowship and European Research Council Consolidator Grant (under the European Union's Horizon 2020 research and innovation program, grant agreement No 723890) at the University of Leicester. GSO and JF were supported by funds from NASA, distributed to the Jet Propulsion Laboratory, California Institute of Technology; JF was supported through JPL's Year-round Internship Program (YIP). This investigation was partially based on thermal-infrared observations acquired at (i) the ESO Very Large Telescope Paranal UT3/Melipal Observatory (program ID 096.C-0091); (ii) Subaru Telescope and obtained from the SMOKA database, which is operated by the Astronomy Data Center, National Astronomical Observatory of Japan (program ID S16B-049); and (iii) NASA's Infrared Telescope Facility, which is operated by the University of Hawaii under contract NNH14CK55B with the National Aeronautics and Space Administration (program ID 2016A-022).


\end{document}